\documentclass [prb,reprint,notitlepage,groupedaddress,floatfix] {revtex4-2}
\usepackage{amsmath}
\usepackage{graphicx}
\usepackage{lmodern}
\usepackage{amssymb}
\usepackage{dsfont}
\usepackage[caption=false]{subfig}
\usepackage[colorlinks=true, citecolor=cyan, urlcolor=red, linkcolor=cyan]{hyperref}
\usepackage[normalem]{ulem}

\renewcommand{\(}{\left (}
\renewcommand{\)} {\right )}
\renewcommand{\]} {\right ]}
\renewcommand{\[} {\left [}

\begin{document}

\title{{Nodal topological superconductivity in nodal-line semimetals}}
\author{Zhenfei Wu}
\author{Yuxuan Wang}
\affiliation{Department of Physics, University of Florida, Gainesville, Florida 32611, USA}
\date{\today}

\begin{abstract}
We analyze possible nodal superconducting phases that emerge from a doped nodal-line semimetal. We show that nodal-line superconducting phases are favored by interactions mediated by  short-range ferromagnetic fluctuations or Hund's coupling. It is found that the leading pairing channels are momentum-independent, orbital-singlet and spin-triplet. In the pairing state, we show that the Bogoliubov-de Gennes (BdG) Hamiltonian hosts a pair of topologically protected nodal rings on the equators of the torus Fermi surface (FS). Using a topological classification for gapless systems with inversion symmetry, we find that these nodal rings are topologically nontrivial and protected by integer-valued monopole charges $\nu = \pm 2$.
In the scenario of pairing driven by  ferromagnetic fluctuations, we analyze the fate of superconductivity in the magnetically ordered phase. Based on Ginzburg-Landau free energy analysis, we find 
the energetically favored superconducting state is characterized by the coexistence of two pairing orders whose $\bf d$-vectors are perpendicular to the magnetization axis $\bf M$ with their phases unfixed. In this case, each nodal loop in the pairing state splits into two, carrying a $\pm 1$ monopole charge. For bulk-boundary correspondence, these nodal rings enclose flat-band Majorana zero modes on top and bottom surface Brillouin Zones with distinct $\mathbb{Z}$-valued topological invariants.
\end{abstract}

\maketitle


\section{Introduction}
Topological semimetals have recently attracted intense research interest in condensed matter physics. These  systems harbor gapless band structures within the three-dimensional (3D) bulk Brillouin Zone (BZ), with a vanishing density of states, and are often topologically protected by specific crystalline symmetries. Depending on the co-dimension of the gapless region, the band crossing can form  either nodal points  \cite{Dirac/weyl,Weyl} or nodal lines \cite{TSM,R_TSM,TSM_BJY,TNLSM,LNSM,NLSM_hyper,TlTaSe2,Bian2016,NRSM2,Chan2016,Dirac_loop,CaAgP,NL_graphene,chain1,chain2,NRSM3,link1,link2,Weyl_link,Dirac_loop,knot}. The line nodes  can form rings \cite{NLSM_hyper,TlTaSe2,Bian2016,NRSM2,Chan2016,Dirac_loop,NL_graphene,chain1,chain2,NRSM3}, chains \cite{chain1,chain2,NRSM3}, links \cite{link1,link2} and other composite structures \cite{knot,Weyl_link,Dirac_loop}. In the absence of spin-orbit coupling, the nodal rings can be further classified into Weyl or Dirac loops depending on the absence or presence of spin degeneracy.

Nodal-line semimetals (NLSMs) are naturally interesting platforms for the interplay between correlation effects and nontrivial topology \cite{Liu2017,Roy2017,Wang_Rahul2017,Wang_Ye2016,Rahul2016,Sur_Rahul2016,Wang_NLSC1}. In particular, many recent theoretical studies have uncovered routes toward novel gapped and nodal topological superconductivity from topological semimetals \cite{Rahul2016,Sur_Rahul2016,Wang_NLSC1,Wang_NLSC2,spin_polarized,Yu_Euler,Wang_Zhou2022,NRSC,Li_Haldane,Lu2015,DSM1,DSM3,DSM2,DSM4}. In doped Weyl semimetals \cite{Li_Haldane,Lu2015} and Dirac semimetals \cite{DSM1,DSM3,DSM2,DSM4}, nodal topological superconducting phases have been studied extensively. In the latter system where crystalline symmetry plays an important role for the normal state topology, it was found that topological nodal pairing also requires crystalline symmetry and only appear for certain unconventional pairing symmetries \cite{DSM2,DSM4}.

Recently, experimental studies have led to the discovery of superconductivity within nodal-loop semimetals \cite{NLSC0,NLSC1,NLSC2,NLSC3,NLSC4,NLSC5,NLSC6,NLSC7,NLSC_comm,Fragile}, while for many of them the pairing symmetry remains to be elucidated.
For such systems, much of the theoretical interest has  focused on fully gapped pairing phases with unconventional pairing symmetry that displays first- and {higher-order topology} protected by crystalline symmetries \cite{Rahul2016,Sur_Rahul2016,Wang_NLSC1,Wang_NLSC2,spin_polarized}. In this work we focus on nodal pairing phases from a doped NLSM with a Dirac loop.  Such Dirac loop structures have been reported in Ca$_3$P$_2$ \cite{Chan2016}, Cu$_3$N \cite{Dirac_loop}, CaAgP and CaAgAs \cite{CaAgP}.

Our key findings are that doped Dirac-loop semimetals host line-nodal pairing phases with $B_{1u}$ pairing symmetry of the $D_{2h}$ group, which are momentum-independent, orbital-singlet and spin-triplet. We study both the pairing mechanism of these orders and their topological classification. We show that short-ranged ferromagnetic fluctuations, as well as Hund's coupling favor these pairing orders as leading superconducting instabilities. The pairing orders are described by a $\bf d$-vector, which is degenerate.
 These pairing orders support a pair of gapless superconducting nodal rings. 
 Despite the similarity to the nodal rings in the normal state, we show that they are characterized by different topological indices. The nodal rings are protected by particle-hole, inversion, and a composite time-reversal symmetry, which is a product of the physical time-reversal and a spin rotation. 
 The Bogoliubov-de Gennes (BdG) Hamiltonian belongs to the CI+$\cal I$ class according to the AZ+$\cal I$ table in Ref. \cite{AZ+I}, which classifies the topological charges of gapless nodes in centrosymmetric systems. By directly computing the topological invariant, we show that the superconducting nodal rings found in this system are protected by nontrivial monopole charges $\nu = \pm 2$.

The topological stability of the nodal rings  can be illustrated by adding symmetry-allowed perturbation terms. In the physical context, we consider the fate of the pairing phases when the ferromagnetic order, whose short-ranged fluctuations mediates superconductivity, becomes long-range. We show via a Ginzburg-Landau analysis that in the presence of a magnetic moment $\bf M$, the leading pairing instability is towards a coexistence of two pairing orders. The two pairing orders belong to the $B_u$ irreducible corepresentation of the magnetic point group No.~8.4.27 \cite{BCS}, and are identical to the pairing orders in the paramagnetic phase with ${\bf d}\perp {\bf M}$ with their relative phase is fixed at $\frac{\pi}{2}$. This can be understood as a fully spin-polarized pairing state on the ``larger" toroidal Fermi surface while the ``smaller" Fermi surface does not favor superconductivity. As the temperature lowers, both Fermi surfaces are gapped (albeit incompletely), and in the BdG spectrum each of the nodal loops in the paramagnetic phase splits into two. To understand their topological properties, we note that while the magnetic order breaks time-reversal symmetry and spin-rotation symmetry, it preserves their product, i.e., the composite time-reversal symmetry that we use for the CI topological classification. Indeed, a direct evaluation of the topological invariant shows that each nodal loop now carries a monopole charge $\nu = \pm 1$. Furthermore, we find that the topological invariant $\nu$ within class CI+$\cal I$ can be interpreted as the difference of the topological invariants of two fully gapped 1D subsystem separated by the superconducting nodal rings. This is demonstrated in the energy spectrum with open boundary condition along $z$ direction, and the corresponding surface Brillouin Zones host flat-band Majorana zero modes enclosed by the superconducting nodal rings.

The rest of this manuscript is organized as follows. In Sec. \ref{sec:normal}, we discuss the normal state Fermi surface of doped NLSMs. In Sec. \ref{sec:fluctuation}, we use Fierz identity to determine that from either ferromagnetic fluctuations or Hund's coupling, the leading pairing channels are both $s$-wave orbital-singlet and spin-triplet pairings. The superconducting critical temperature is also derived for these pairing channels. In Sec. \ref{sec:projection}, we show that these pairings get projected onto the torus Fermi surfaces  and exhibit a pair of nodal rings on the equator due to the nontrivial pseudospin textures. In Sec. \ref{sec:ferro}, we investigate the leading pairing channel in the presence of a ferromagnetic order using a Ginzburg-Landau free energy analysis. In Sec. \ref{sec:winding}, we analyze the topological protection of nodal-ring superconducting orders in both paramagnetic and ferromagnetic phases as well as the bulk-boundary correspondence and summarize our results  in Sec. \ref{sec:conclusion}.

\section{Lattice model}
\label{sec:normal}

\begin{figure}
    \includegraphics[width = 2in]{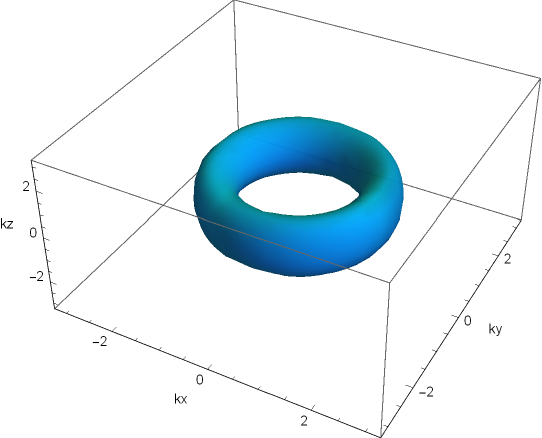}\\
    \caption{A schematic torus-like Fermi surface from $H_0(\mathbf{k})$ with parameters $t_1 = 2.46, t_2 = 0.5$ and $\ \mu = 1.2$.}
    \label{fig:FS}
\end{figure}

A lattice model Hamiltonian for a nodal-ring semimetal is given by \cite{Wang_NLSC2}
\begin{align}
	H_0({\bf k})  =\ & (6 - t_1 - 2\cos k_x -2 \cos k_y - 2\cos k_z)\sigma_z s_0 \nonumber\\
	 & + 2t_2\sin k_z \sigma_x s_0 - \mu \sigma_0 s_0,
	 \label{eq:H0}
\end{align}
where $\sigma_i (s_i)$ denotes the $i$-th Pauli matrix representing the orbital(spin) degrees of freedom and the implicit tensor product (i.e., $\sigma_0\otimes s_0$) is assumed. In Eq. \eqref{eq:H0}, ($6-t_1$) represents the difference of on-site energies between two orbitals and the momentum-dependent part of the first term denotes the difference of nearest-neighbor intra-orbital hoppings. The second term in Eq. \eqref{eq:H0} is the nearest-neighbor inter-orbital hopping along $\hat{z}$ direction and $\mu$ is the chemical potential. Here we neglect the term $\epsilon({\bf k})\sigma_0s_0$ in the dispersion which does not influence the normal state topology. For $0 < t_1 < 4$, $H_0(\bf k)$ displays a nodal ring ($k_z = 0, \cos k_x + \cos k_y = 2-t_1/2$) in 3D Brillouin Zone at half filling ($\mu = 0$). When $t_1$ is continuously tuned from positive to negative, the nodal ring shrinks to a point and annihilates itself. For finite but small doping ($|\mu| < t_1$), the nodal ring  inflates into a torus-like Fermi surface (FS),   shown in Fig. \ref{fig:FS}. Every point on the FS is two-fold degenerate.

The Hamiltonian in Eq.~\eqref{eq:H0} preserves inversion and time-reversal symmetry
\begin{align}
    \hat{\cal I} H_0({\bf k}) \hat{\cal I}^{-1} &= H_0(-{\bf k}),   \\
    \hat{\cal T} H_0({\bf k}) \hat{\cal T}^{-1} &= H_0(-{\bf k}), 
\end{align}
where $\hat{\cal I} = \sigma_z$ and $\hat{\cal T} = \sigma_z {\cal K}$ (${\cal K}$ is the complex conjugate operator). $H_0({\bf k})$ also preserves SU(2) spin-rotation symmetry due to the absence of spin-orbit coupling, and hence $\hat{\cal T}^2 = +1$. Correspondingly, $H_0({\bf k})$ belongs to class AI$+\cal I$ \cite{AZ+I} and the nodal ring is robust against symmetry-preserved perturbations due to the Berry phase $\pi$ of a Wilson loop which interlocks with the nodal ring \cite{Fang2015}. {Different from a Weyl loop, due to time-reversal symmetry, here the nodal-ring is four-fold degenerate and is dubbed a ``Dirac loop" \cite{Chan2016,Dirac_loop,CaAgP,NL_graphene,chain1,chain2}.}

\section{Pairing Mechanism}
\label{sec:fluctuation}

In this section, we analyze superconducting instabilities mediated by two similar types of interactions: short-range ferromagnetic fluctuations and the inter-orbital Hund's coupling. We find that both pairing mechanisms favor $s$-wave, orbital singlet and spin-triplet channels. In Sec. \ref{sec:projection}, we show that these pairing channels exhibit a pair of superconducting nodal rings which is attributed to the nontrivial FS pseudo-spin textures.

\subsection{Ferromagnetic fluctuations}
\label{subsec:ferro}

\begin{figure}
	\subfloat[]{\includegraphics[width=0.45\columnwidth]{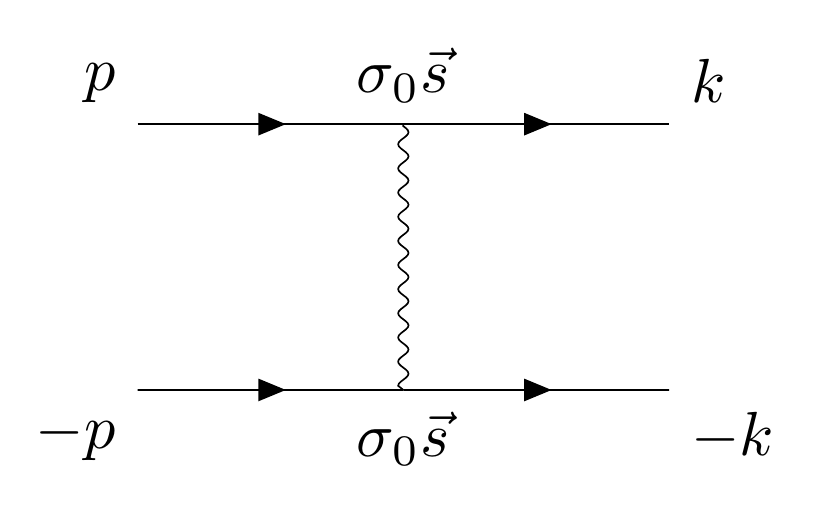}}
	\quad
	\subfloat[]{\includegraphics[width=0.45\columnwidth]{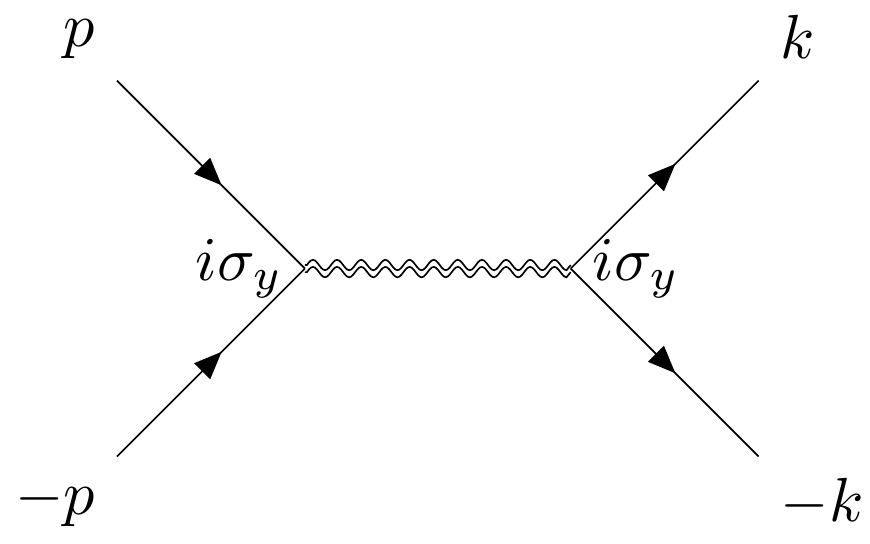}}
   \caption{Diagrammatic representations of (a) ferromagnetic spin fluctuations and (b) ferromagnetic fluctuations converted to Cooper pairing channels. Solid lines represent fermionic propagators.  The single-wavy line in (a) denotes the interaction $V_0$ while the double wavy-line in (b) denotes the interaction $V_0/4$ after applying the Fierz identity. The vertices in (b) contain a spin-triplet part ($\vec{s}\ is_y$) which is not shown in the figure.} 
	\label{fig:ferro}
\end{figure}

We consider a short-range ferromagnetic fluctuation among all orbitals  (diagrammatically shown in Fig. \hyperref[fig:ferro]{\ref{fig:ferro}(a)}), 
\begin{align}
    H_{\text{ferro}} =  V_0\int d{\bf p}d{\bf k} c^{\dag}({\bf p})\sigma_0\vec{s}\ c({\bf k}) \cdot
    c^{\dag}(-{\bf p})\sigma_0\vec{s}\ c(-{\bf k}),
    \label{eq:pairing}
\end{align}
where $c^{\dag}({\bf p}) = [\psi^{\dag}_{\bf p,+,\uparrow},\ \psi^{\dag}_{\bf p,+,\downarrow},\ \psi^{\dag}_{\bf p,-,\uparrow},\ \psi^{\dag}_{\bf p,-,\downarrow}]$ 
is a four-component fermionic creation operator and $V_0 < 0$. Here $\pm$ represents the orbital degree of freedom and $\uparrow\downarrow$ labels the spin. 
This interaction can be decomposed into different orbital and spin pairing channels separately by means of the Fierz identity \cite{Fierz,Lin2018} (also see Appendix \ref{appen:1}) and we find
\begin{align}
    &\ c^{\dag}({\bf p})\sigma_0\vec{s}\ c({\bf k}) \cdot c^{\dag}(-{\bf p})\sigma_0\vec{s}\ c(-{\bf k})\nonumber\\
    = &\ \frac{1}{4}\sum_{\substack{a = 0,x,y,z\\ b = x,y,z}}c^{\dag}({\bf p})\sigma_a i\sigma_y\otimes s_b is_y\[c^{\dag}(-{\bf p})\]^{\text{T}}\nonumber\\
    &\hspace{.8in} \times \[c(-{\bf k})\]^{\text{T}}\(\sigma_a i\sigma_y\)^{\dag}\otimes \(s_b is_y\)^{\dag} c({\bf k}).
    \label{eq:decomp}
\end{align}
The result in Eq. \eqref{eq:decomp} shows that ferromagnetic fluctuations naturally favor spin-triplet pairings \cite{spin_polarized,ferro_triplet,near_ferro_triplet}.  Note that we approximately set the scattering amplitude in Fig. \hyperref[fig:ferro]{\ref{fig:ferro}(a)} to be a constant $V_0$ which does not contribute to momentum transfer due to the short-range behavior of the ferromagnetic fluctuation. Moreover, the Pauli exclusion principle imposes constraints on the pairing function, $\Delta_{\bf k} = -\Delta_{-\bf k}^{\text{T}}$. Hence the leading pairing channel is expected to be momentum-independent, orbital-singlet and spin-triplet. After neglecting the orbital-triplet channels in Eq. \eqref{eq:decomp}, the interaction in Eq. \eqref{eq:pairing} can be rewritten as 
\begin{align}
    H_{\text{int}} &= \frac{V_0}{4} \int d{\bf p}d{\bf k} c^{\dag}({\bf p}) i\sigma_y\otimes \vec{s}\ is_y\[c^{\dag}(-{\bf p})\]^{\text{T}} \nonumber\\
     &\hspace{1in} \cdot  \[c(-{\bf k})\]^{\text{T}}(i\sigma_y\otimes \vec{s}\ is_y)^{\dag} c({\bf k}),
     \label{eq:Hint}
\end{align}
diagrammatically shown in Fig. \hyperref[fig:ferro]{\ref{fig:ferro}(b)}.

The critical temperature of these orbital-singlet and spin-triplet pairing channels in the paramagnetic phase can be derived from a normal state FS instability, which is captured by the linearized gap equation \cite{Atland_Simons}
\begin{align}
1 = -\frac{V_0T}{4}  \sum_k \text{Tr}\[\sigma_y s_j is_y G_0(k) \sigma_y (s_j is_y)^{\dag} G_0^{\text{T}}(-k)\],
    \label{eq:bubble}
\end{align}
where $k \equiv ({\bf k},\omega_n)$ and $\omega_n = (2n + 1)\pi T$ is the fermionic Matsubara frequency. $G_0(k)$ is the normal state Green's function $G_0(k) = \[i\omega_n - H_0({\bf k})\]^{-1}$ and $j = x/y/z$ denotes three distinct spin-triplet pairing channels. The critical temperatures for three spin-triplet channels are the same by SU(2) spin-rotation symmetry, hence we only consider one particular spin index (e.g., $s_x$) throughout the rest of this subsection.  In order to explicitly evaluate the critical temperature, we write down a $\bf k\cdot p$ continuum model for a nodal-ring semimetal \cite{TlTaSe2,Bian2016,Li2018} from Eq. \eqref{eq:H0} 
\begin{align}
    H_0({\bf k}) = \(\frac{k_x^2 +k_y^2}{m^*} - t_1\)\sigma_z + v_z k_z \sigma_x - \mu,
    \label{eq:kdotp}
\end{align}
where $m^*$ is the effective mass and $v_z$ is the Fermi velocity along $z$ direction. To further simplify the notation, we set $k_p \equiv (k_x^2 + k_y^2)/m^* - t_1$ and replace $k_z \to v_zk_z$ for now ($v_z$ is resumed in the expression of density of states). The normal state Green's function is
\begin{align}
    G_0(k) = \frac{(i\omega_n + \mu)\sigma_0 + k_p \sigma_z + k_z \sigma_x}{(i\omega_n + \mu)^2 - k_p^2 - k_z^2}.
\end{align}
The integrand in Eq. \eqref{eq:bubble} is simplified to
\begin{align}
    &\ \text{Tr}\[\sigma_ys_x G_0(k) \sigma_ys_x G_0^{\text{T}}(-k)\] \nonumber\\
      =&\ 2\text{Tr}\Bigg[\frac{(i\omega_n + \mu)\sigma_0 - k_p \sigma_z - k_z \sigma_x}{(i\omega_n+\mu)^2 - k_r^2}\nonumber\\
    & \hspace{1.2in} \times \frac{(-i\omega_n + \mu)\sigma_0 + k_p \sigma_z - k_z \sigma_x}{(-i\omega_n+\mu)^2 - k_r^2}\Bigg]\nonumber\\
    =&\ \frac{4(\omega_n^2 + \mu^2 - k_p^2 + k_z^2)}{\[(i\omega_n+\mu)^2 - k_r^2\]\[(-i\omega_n+\mu)^2 - k_r^2\]},
\end{align}
where $k_r \equiv \sqrt{k_p^2 + k_z^2}$ and the frequency summation in Eq. \eqref{eq:bubble} yields
\begin{align}
     &\ T\sum_{n} \text{Tr}\[\sigma_ys_x G_0(k) \sigma_ys_x G_0^{\text{T}}(-k)\] \nonumber\\
   =&\ 4\int_{C} \frac{dz}{2\pi i} f(z) \frac{-z^2 + \mu^2}{\[(z+\mu)^2 - k_r^2\]\[(-z+\mu)^2 - k_r^2\]},
\end{align}
where $f(z) = 1/(e^{\beta z} + 1)$. There are four roots in the denominator $z_{1,2} = \pm k_r -\mu$, $z_{3,4} = \pm k_r + \mu$.  Note that the momentum integral $\sum_{\bf k} \equiv \frac{m^*}{8\pi^2v_z} \int_{\bf k} dk_p dk_z$ is performed within a narrow region around the Fermi surface ($k_r = \mu$), which is even for both $k_p$ and $k_z$. Assume an energy cutoff $\omega_c$ around FS, we obtain $z_1 \in [-\omega_c, \omega_c]$ and $z_2 \simeq -2\mu$. After applying residue theorem, Eq. \eqref{eq:bubble} becomes 
\begin{align}
       1 &  \simeq -\frac{V_0}{4}\sum_{\bf k}\(\frac{1}{2}\frac{\tanh \frac{\beta z_1}{2}}{z_1} + \frac{3}{2}\frac{\tanh \frac{\beta z_2}{2}}{z_2} \) \nonumber\\
      &\simeq -\frac{V_0}{4}  \frac{N(0)}{2} \int_{-\omega_c}^{\omega_c} d\epsilon \frac{\tanh \frac{\beta\epsilon}{2}}{\epsilon} \nonumber\\
      &\simeq -\frac{V_0}{4}  N(0) \log{\frac{\omega_c}{T}},
\end{align}
where the density of states at Fermi energy $N(0)$ is derived by noting that the total number of states below the FS is 
\begin{equation}
    \frac{m^*}{8\pi^2 v_z}\int_{k_r\leq \mu}dk_p dk_z = \frac{m^*}{8\pi^2 v_z}\pi \mu^2 = \frac{m^*\mu^2}{8\pi v_z},
\end{equation}
hence
\begin{equation}
    N(0) = \frac{d}{d\mu} \frac{m^*\mu^2}{8\pi v_z} = \frac{m^*|\mu|}{4\pi v_z}.
    \label{eq:dos}
\end{equation}
Together with the linearized gap equation, the critical temperature is found to be 
\begin{align}
     T_c = \omega_c \exp \(-\frac{16\pi v_z}{m^*|\mu V_0|}\),
\label{eq:Tc}
\end{align}
It is important to emphasize that our theory is within the weak-pairing regime, which is inapplicable to the half-filling case ($\mu = 0$) where the density of states vanishes. As a consequence, a strong pairing mechanism is necessary to ensure  a superconducting instability at half-filling, which is beyond the scope of our work.

\subsection{Hund's coupling}
\begin{figure}
	\includegraphics[width=0.8\columnwidth]{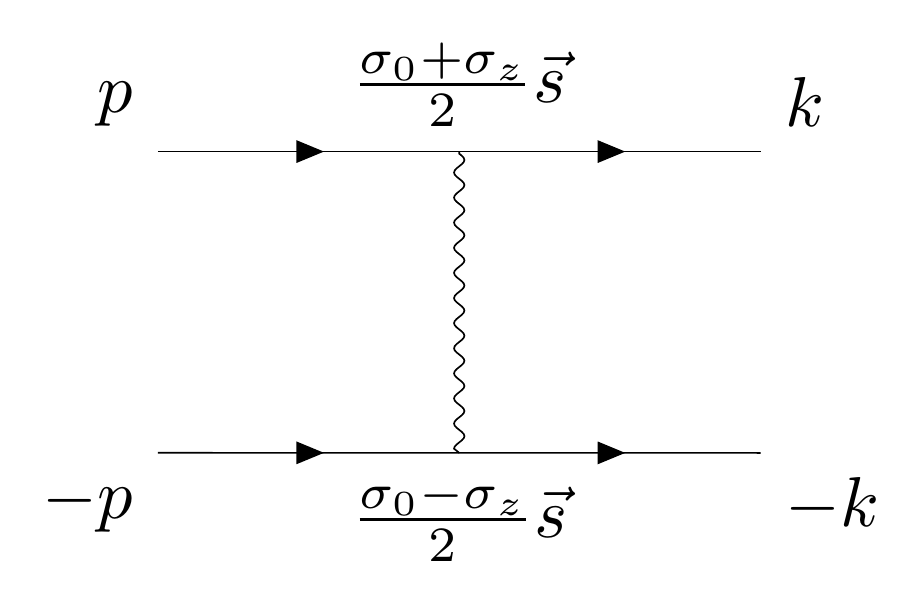}
   \caption{Diagrammatic representation of Hund's coupling.} 
\label{fig:hunds}
\end{figure}

In this subsection, we analyze the pairing orders from Hund's coupling, which is an effective local ferromagnetic coupling between different orbitals \cite{Hunds,Lee_Wen08}. Following the Feynman diagram in Fig. \ref{fig:hunds}, the Hund's coupling is expressed as
\begin{align}
    H_{\text{Hunds}} =  &\ V_H\int d{\bf p}d{\bf k} c^{\dag}({\bf p})\frac{\sigma_0 +\sigma_z}{2}\otimes\vec{s}\ c({\bf k}) \nonumber\\
    &\hspace{.7in}  \cdot c^{\dag}(-{\bf p})\frac{\sigma_0 - \sigma_z}{2}\otimes\vec{s}\ c(-{\bf k}), 
    \label{eq:hunds}
\end{align}
where $V_H < 0$. Compared with Eq. \eqref{eq:decomp}, the spin channel decomposition from Fierz identity is the same as ferromagnetic fluctuations, both being spin-triplet. Nevertheless, the orbital part decomposition is different (details in Appendix \ref{appen:1}), 
\begin{align}
    &\ c^{\dag}({\bf p})\frac{(\sigma_0 + \sigma_z)}{2}\otimes\vec{s}\ c({\bf k}) \cdot c^{\dag}(-{\bf p})\frac{(\sigma_0 - \sigma_z)}{2}\otimes\vec{s}\ c(-{\bf k})\nonumber\\
    =&\ \frac{1}{8}\sum_{\substack{\{i,j\}\\ a = x,y,z}}c^{\dag}({\bf p})\sigma_i i\sigma_y \otimes s_a is_y\[c^{\dag}(-{\bf p})\]^{\text{T}} \nonumber\\
    &\hspace{.75in}  \times \[c(-{\bf k})\]^{\text{T}}\(\sigma_j i\sigma_y\)^{\dag}\otimes \(s_a is_y\)^{\dag} c({\bf k}),
    \label{eq:decomp2}
\end{align}
where the summation of indices $\{i,j\}$ runs over four different combinations $\{0,0\},\{z,0\},\{0,z\},\{z,z\}$ such that the orbital parts of scattering vertices in Eq. \eqref{eq:decomp2} are $\{i\sigma_y, i\sigma_y\}$, $\{\sigma_x, i\sigma_y\}$, $\{i\sigma_y, \sigma_x\}$ and $\{\sigma_x, \sigma_x\}$. Similar to the case of ferromagnetic fluctuations, Hund's coupling is also an effective short-range interaction, thus the pairing function should be momentum-independent. Accordingly, the only possible orbital part decomposition in Eq. \eqref{eq:decomp2}  is $\{i\sigma_y, i\sigma_y\}$ in compliance with the Pauli exclusion principle. The pairing interaction from Hund's coupling has the same form as Eq. \eqref{eq:Hint}, which confirms that $s$-wave orbital-singlet and spin-triplet pairing channels are also attractive mediated from Hund's coupling. The derivation of the critical temperature is similar, and one only needs to replace $V_0$ by $V_H/2$ in Eq. \eqref{eq:Tc}, 
\begin{align}
	T_c = \omega_c \exp \(-\frac{32\pi v_z}{m^*|\mu V_H|}\).
\label{eq:Tc_Hunds}
\end{align}

\subsection{Other interactions}

Apart from the attractive interactions mentioned above, we also investigate the effects from repulsive Hubbard interactions  
\begin{align}
	H_\text{Hubbard} =  U \sum_{i,\sigma} n_{i\sigma\uparrow}n_{i\sigma\downarrow} + U'\sum_{\substack{i,s,s'\\ \sigma \neq\sigma'}} n_{i\sigma s}n_{i\sigma' s'},
\end{align}
where $U, U' > 0$ denote on-site and extended Hubbard repulsions, $i$ is unit cell index in real space and $n_{i\sigma s}\equiv c_{i\sigma s}^{\dag}c_{i\sigma s}$ is the density operator. We find that the on-site intra-orbital repulsion $U$ has no corrections to the orbital-singlet and spin-triplet channels. On the other hand, the extended term $U'$ has a destructive contribution to the orbital-singlet channels. Given our focus on the scenarios where ferromagnetic fluctuations or Hund's coupling becomes dominant, we can justifiably neglect the contributions from the sub-dominant extended Hubbard term.

\section{Nodal-ring superconductivity}
\label{sec:projection}
\begin{figure}
    \centering
    \includegraphics[width = 3in]{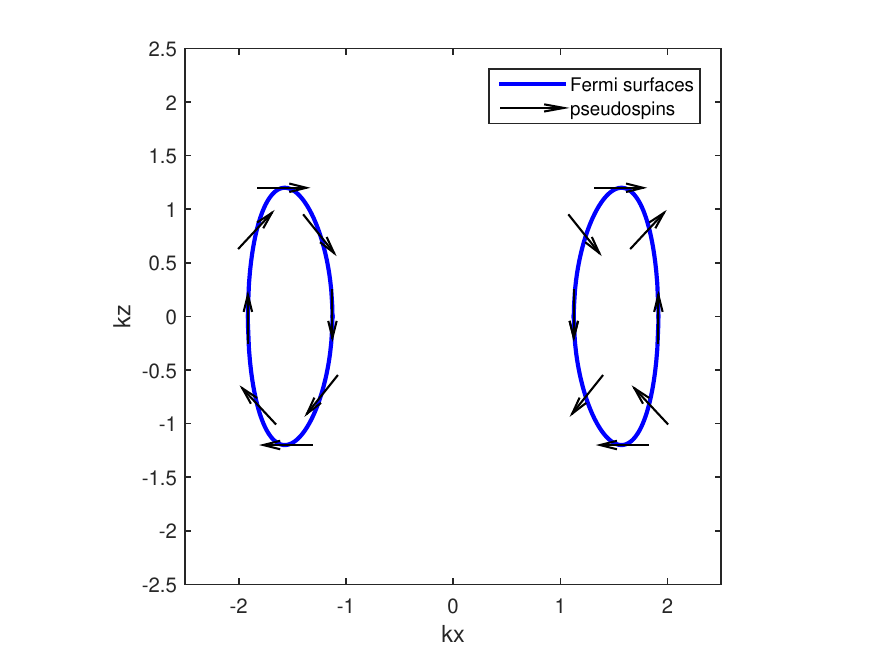}\\
    \caption{Pseudo-spin textures on FS at $k_y = 0$ with parameters $t_1 = 2.46$, $t_2 = 0.5$ and $\mu = 1.2$. The blue curves represent the cross sections of FS on the
    $k_y = 0$ plane and the black arrows denote pseudo-spin orientations.}
    \label{fig:projection}
\end{figure}

\begin{figure}
    \includegraphics[width = 2in]{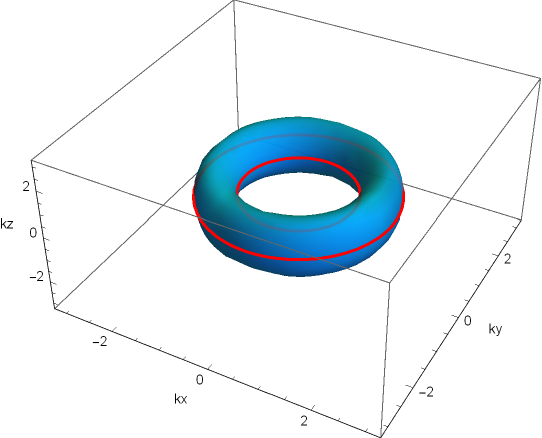}\\
    \caption{Two superconducting nodal rings (red) on the equators of the torus Fermi surface.}
    \label{fig:SC_nodal_ring}
\end{figure}

In this section, we show that $s$-wave orbital-singlet and spin-triplet pairing orders exhibit a nodal gap structure on the equators of the torus FS. In order to verify the gap nodes, we project the pairing orders onto the torus FS since the pairing instability comes from electronic states near FS \cite{Fierz}. We note that the periodic parts of the Bloch wavefunction  $|\pm,\bf k\rangle$ of the normal state Hamiltonian $H_0(\mathbf{k})$ in Eq. \eqref{eq:kdotp} are given by
\begin{align}
    H_0({\bf k})|\pm,{\bf k}\rangle = \varepsilon_{\pm,\bf k} |\pm,\bf k\rangle,
\end{align}
with energies $\varepsilon_{\pm,\bf k} = \pm \sqrt{k_p^2 + k_z ^2} - \mu$. Without loss of generality, we assume a positive and small $\mu$ which satisfies $0 < \mu < t_1$. The Fermi surface is a torus given by $k_p^2 + k_z^2 = \mu^2$ and the Bloch state on FS is 
\begin{align}
    |+,{\bf k}\rangle = 
    \frac{1}{\sqrt{2\mu(\mu - k_p)}}
    \[\begin{array}{c}
        -k_z\\
        k_p - \mu 
    \end{array}\].
\end{align}
In Fig. \ref{fig:projection}, we plot orbital pseudo-spin textures on the FS contour at $k_y = 0$ (real spins lack nontrivial polarizations so we suppress them). The orbital-singlet pairing order $c_{\bf k}^{\dag}\(i\Delta \sigma_y\) \(c_{-\bf k}^{\dag}\)^{\text{T}}$ considered in Eq. \eqref{eq:Hint} can be projected onto the FS as \cite{Fierz}
\begin{align}
    \Delta^{\text{FS}}({\bf k}) = \langle +,{\bf k}| i\Delta \sigma_y \(|+,-{\bf k}\rangle^*\) = \frac{\Delta k_z}{\mu}, 
    \label{eq:projected}
\end{align}
which exhibits two superconducting nodal rings located at $k_z = 0$. The vanishing  pairing amplitudes on the equators can also be deduced from the orbital pseudo-spin textures on FS shown in Fig. \ref{fig:projection}:
$i\Delta \sigma_y$ is a pseudo-spin singlet state, while the electronic states with opposite momentum at $k_z = 0$ possess the same pseudo-spin polarizations. Therefore,  electrons at $k_z = 0$ cannot form Cooper pairs in the orbital-singlet channels. This leads to a pair of superconducting nodal rings at the equators of the torus Fermi surface (see Fig. \ref{fig:SC_nodal_ring}). The aforementioned nodal-ring superconductivity is a common feature for all three spin-triplet pairing channels in the paramagnetic phase.

\section{Fate of superconductivity in the ferromagnetic phase}
\label{sec:ferro}

When ferromagnetic fluctuations become long-ranged, the system may develop a ferromagnetic order below the Curie temperature. Here we investigate the fate of pairing orders in the presence of ferromagnetism. The normal state Hamiltonian becomes 
\begin{align}
    H_0'({\bf k})  =\ & (6 - t_1 - 2\cos k_x -2 \cos k_y - 2\cos k_z)\sigma_z s_0 \nonumber\\
	 & + 2t_2\sin k_z \sigma_x s_0 - \mu \sigma_0 s_0 - M_z \sigma_0 s_z,
	 \label{eq:ferro_normal}
\end{align}
where we assume the magnetization along $z$ axis. The Fermi surface is  split into two spin-polarized sectors by the magnetic order \cite{Chan2016}. The interplay between the ferromagnetic order and spin-triplet superconductivity depends on the relative orientations between the magnetization and $\bf d$-vectors of superconducting order parameters. The matrix forms of order parameters describing ferromagnetism and superconductivity are $\sigma_0 ({\bf M}\cdot \bf s)$ and $i\sigma_y ({\bf d}\cdot{\bf s}) is_y$ (we are using the conventional definition of $\bf d$-vector for spin-triplet $p$-wave pairings here, although the $\bf d$-vector is independent of $\bf k$).  To be specific, we denote the orbital-singlet and  spin-triplet pairing orders by 
\begin{align}
	\Delta_x i\sigma_y s_xis_y\qquad \text{for}\quad d_x\quad \text{component},\nonumber\\
	\Delta_y i\sigma_y s_yis_y\qquad \text{for}\quad d_y\quad \text{component},\nonumber\\
	\Delta_z i\sigma_y s_zis_y\qquad \text{for}\quad d_z\quad \text{component}.
\end{align}

In order to determine the favored pairing channel, we perform a Ginzburg-Landau (GL) free energy analysis. 
Without loss of generality, we choose ${\bf M} = (0,0,M_z)$. Including the {quadratic order} term in $\bf d$ and the lowest-order coupling between superconducting orders and {a pre-formed}  ferromagnetic order $\bf M$, the GL free energy can be generally expressed as \cite{GL_ferroSC}
\begin{align}
F({\bf d},{\bf d}^*) = \alpha (T - T_c) {\bf d}\cdot{\bf d}^* + i\gamma {\bf M}\cdot({\bf d}\times {\bf d}^*),
            \label{eq:GL}
\end{align}
where $\alpha > 0$ and $T_c$ is the critical temperature given in Eq. \eqref{eq:Tc}. The three spin-triplet orders share the same $T_c$ in the absence of ferromagnetism. When $T < T_c$, the negative prefactor of the first term supports a finite order parameter $|{\bf d}| \neq 0$. The coefficient $\gamma$ in Eq. \eqref{eq:GL}  can be evaluated {from the Feynman diagram calculation in Appendix \ref{appen:diagram}, giving rise to $\gamma = -N(0)/\mu$.} Since $\gamma< 0$, the leading pairing channel has a relative phase $\pi/2$ between $d_x$ and $d_y$ components. This can be verified from the second term in Eq. \eqref{eq:GL}
\begin{align}
	i\gamma {\bf M}\cdot({\bf d}\times {\bf d}^*) & = i\gamma M_z(d_x d_y^* - d_y d_x^*) \nonumber\\
	& = \gamma M_z |d_x||d_y| \( ie^{i(\alpha_x - \alpha_y)} - ie^{i(\alpha_y-\alpha_x)}\) \nonumber\\
	& = 2\gamma M_z |d_x||d_y| \sin (\alpha_y - \alpha_x),
	\label{eq:gamma2}
\end{align}
where $\alpha_x(\alpha_y)$ is the phase factor carried by the $d_x(d_y)$ component. In order to minimize the free energy, we obtain $\alpha_y - \alpha_x = \pi/2$.
Therefore, ${\bf d}\propto (1,i,0)$ and the Cooper pair carrys a $z$-component total spin $S_z = +1$. To understand this, we recall that  ${\bf m} = i{\bf d} \times {\bf d}^*$ is the magnetic moment of the Cooper pair in spin-triplet pairing channels. The second term in Eq. \eqref{eq:GL} is nothing but the potential energy of a magnetic dipole placed in an external magnetic field $\bf M$. The spin polarizations of the Cooper pair and the ferromagnetic order are aligned with each other so as to minimize the potential energy. This pairing state is an analogy of the superfluid $^3$He-$A1$ phase.

As the temperature is further lowered, the emergence of a sub-dominant pairing channel is anticipated. The secondary transition depends on the quartic order terms: (${\bf d}\cdot{\bf d}^*)^2, |{\bf d}\cdot{\bf d}|^2$ and $({\bf d}\times{\bf d}^*)^2$, which are not included in Eq. \eqref{eq:GL}. With the primary pairing channel already identified, it is more straightforward to evaluate the quartic order terms in the free energy from the following transformations
\begin{align}
\Delta_a = \ d_x - i d_y, \quad \Delta_b = &\ d_x + i d_y, \quad \Delta_z = d_z,
\end{align}
where $\Delta_a$ is the amplitude of $|\uparrow\uparrow\rangle$ spin pairing {which denotes the primary channel} and $\Delta_b$ denotes $|\downarrow\downarrow\rangle$ spin pairing. The GL free energy is
{\begin{align}
F(\Delta_a,\Delta_b,\Delta_z) = &\  -\alpha' \(|\Delta_a|^2 + |\Delta_b|^2 + 2|\Delta_z|^2\) \nonumber\\
& - \frac{|\gamma| M_z}{2} \(|\Delta_a|^2 - |\Delta_b|^2\) \nonumber\\
& + \beta_a |\Delta_a|^4  + 4\tilde{\beta}|\Delta_a|^2|\Delta_z|^2,
            \label{eq:GL2}
\end{align}
where $\alpha' \equiv -\alpha(T-T_c)/2 > 0$ and $\beta_a = \tilde{\beta} = \beta/4 = N(0)/(16\pi^2 T^2)$ (see details in Appendix \ref{appen:diagram}). We have included only quadratic order terms in $\Delta_b$ and $\Delta_z$ in the free energy above, which are sufficient to determine the secondary phase transition. Note that the term $|\Delta_a|^2|\Delta_b|^2$ does not show up in the free energy because it couples fermions from different spin sectors. In order to determine the secondary pairing channel, one needs to check the sign change of two quadratic order terms $|\Delta_b|^2$ and $|\Delta_z|^2$. Based on this reasoning, we set 
\begin{align}
	-\alpha' + \frac{|\gamma| M_z}{2} = 0,
\end{align}
which is the condition for the prefactor of $|\Delta_b|^2$ to vanish. The free energy in Eq. \eqref{eq:GL2} becomes
\begin{align}
	F(\Delta_a,\Delta_z)  =& \ -|\gamma|M_z|\Delta_a|^2 + \frac{\beta}{4}|\Delta_a|^4 \nonumber\\
	  &  +\(-|\gamma|M_z + \beta |\Delta_a|^2\)|\Delta_z|^2.
\end{align}
The magnitude of the primary order $|\Delta_a|$ can be determined by setting $\partial F/\partial \Delta_a = 0$, which yields $|\Delta_a|^2 = 2|\gamma|M_z/\beta$. Therefore, the prefactor of $|\Delta_z|^2$ term is 
\begin{align}
	-|\gamma|M_z + \beta |\Delta_a|^2 = |\gamma|M_z > 0, 
\end{align}
which indicates that $\Delta_z$ has not developed yet. Therefore the $\Delta_b$ pairing channel is favored compared with $\Delta_z$. } 

For a finite magnetic order $\bf M$, the free energy analysis fails because higher-order terms in $\bf M$ (e.g., ${\bf M}^2, {\bf M}^3, \cdots$) are not negligible. In general, one needs a non-perturbative method to analyze the interplay between ferromagnetism and superconductivity. Nevertheless, we argue here that the story is qualitatively the same as that for a small ${\bf M}$, i.e., $\Delta_a$ is the primary order and $\Delta_b$ is secondary while $\Delta_z$ is disfavored. The underlying reason is as follows: $\Delta_z \sigma_y s_x$ pairs electrons from two Fermi surfaces with opposite spins, which is negligible compared with intra-FS equal-spin pairing terms in the weak pairing regime. Also, the FS with $|\uparrow\uparrow\rangle$ spin polarization has a greater density of states, thus favoring $\Delta_a$ compared with $\Delta_b$.

\section{Topology of the nodal-ring superconducting orders}
\label{sec:winding}
\subsection{Paramagnetic phase}

To analyze the topological properties of the aforementioned $s$-wave orbital-singlet and spin-triplet pairing channels, we write down the Bogoliubov-de Gennes (BdG) Hamiltonian of the superconducting nodal-ring system in the mean field regime
\begin{align}
    {\cal H}_{\text{BdG}}({\bf k})= \[\begin{array}{cc}
        H_0({\bf k}) & -\vec{\Delta}\cdot i\sigma_y\vec{s}\ is_y\\
        -\vec{\Delta}^{\dag}\cdot(i\sigma_y\vec{s}\ is_y)^{\dag} & -H_0^{\text{T}}(-{\bf k})
    \end{array}
    \],
\end{align}
where the second quantized Hamiltonian is  ${\cal H} = \frac12\int_{\bf k}\Psi^{\dag}_{\bf k} {\cal H}_{\text{BdG}}({\bf k})\Psi_{\bf k}$ and  the Nambu spinor is defined as  $\Psi_{\bf k} = \[c^{\text{T}}({\bf k}),\  c^{\dag}(-\bf k)\]^{\text{T}}$. 

In the paramagnetic phase, due to the spin degeneracy, the BdG Hamiltonian can be decoupled into two identical copies of four-band models, each given by
\begin{align}
    {\cal H}_{\text{BdG}}({\bf k}) = &\ (6 - t_1 -2 \cos k_x - 2\cos k_y - 2\cos k_z) \sigma_z\tau_z \nonumber\\
    &+ 2t_2\sin k_z \sigma_x\tau_0  - \mu\sigma_0 \tau_z + \Delta \sigma_y\tau_y,
    \label{eq:BdG}
\end{align}
where  $\tau_{i}$ is the $i$-th Pauli matrix in the Nambu space and $\tau_0$ is the identity.  ${\cal H}_{\text{BdG}}({\bf k})$ preserves inversion and time reversal symmetry, and additionally,  particle-hole symmetry ($\hat{\cal P}$) and chiral symmetry ($\hat{\cal S}$)
\begin{align}
    \hat{\cal I}{\cal H}_{\text{BdG}}({\bf k})\hat{\cal I}^{-1} &= {\cal H}_{\text{BdG}}(-{\bf k}),  \nonumber \\
    \hat{\cal T}{\cal H}_{\text{BdG}}({\bf k})\hat{\cal T}^{-1} &= {\cal H}_{\text{BdG}}(-{\bf k}), \nonumber\\
    \hat{\cal P}{\cal H}_{\text{BdG}}({\bf k})\hat{\cal P}^{-1} &= -{\cal H}_{\text{BdG}}(-{\bf k}), \nonumber\\
    \hat{\cal S}{\cal H}_{\text{BdG}}({\bf k})\hat{\cal S}^{-1} &= -{\cal H}_{\text{BdG}}({\bf k}) , 
    \label{eq:symm}
\end{align}
where $\hat{\cal I} = \sigma_z\tau_z$, $\hat{\cal T} =  \sigma_z \tau_z \cal K$, $\hat{\cal P} = \tau_x\cal K$ and $\hat{\cal S} \equiv i\hat{\cal P}\hat{\cal T} = \sigma_z\tau_y$. Moreover,  Eq. \eqref{eq:BdG} preserves three mirror symmetries $\hat{\cal{M}}_x = \mathds{1}$, $\hat{\cal M}_y = \mathds{1}$ and $\hat{\cal M}_z = \sigma_z\tau_z$, which characterizes the Hamiltonian by $D_{2h}$ point group symmetry. The corresponding pairing function $\Delta \sigma_y \vec{s}\ is_y$ with three distinct $\bf d$-vectors all belong to the $B_{1u}$ irreducible representation \cite{spin_polarized}.

The BdG quasiparticle spectrum can be directly solved from the Hamiltonian in Eq. \eqref{eq:BdG}
\begin{widetext}
\begin{align}
    E({\bf k}) = \pm \sqrt{f^2({\bf k}) + 4t_2^2\sin^2 k_z + \mu^2 + \Delta^2 \pm 2\sqrt{f^2({\bf k})\(\mu^2 + \Delta^2\)  + 4\mu^2 t_2^2\sin^2 k_z}},
\end{align}
\end{widetext}
where $f({\bf k}) = (6 - t_1 -2 \cos k_x - 2\cos k_y - 2\cos k_z)$. By setting $E({\bf k}) = 0$, a pair of gapless nodal rings can be found at $k_z = 0, \cos k_x +\cos k_y = 2-t_1/2 \pm \sqrt{\mu^2+\Delta^2}/2  \simeq 2-t_1/2 \pm \mu$, which resides at two equators of the torus Fermi surface ($\simeq$ comes from weak pairing assumption $\Delta \ll \mu$). This result is consistent with the projected gap found in Eq. \eqref{eq:projected}, where we find the gap vanishing at $k_z = 0$.

The two superconducting nodal rings are found to be robust against perturbations that preserve the symmetries listed in Eq. \eqref{eq:symm}. Therefore, these gapless rings must carry some non-trivial topological charges which prevent them from being gapped. To check the robustness of the two nodal rings, we first notice that  $(\hat{\cal T}\hat{\cal I})^2 = +1$ and $(\hat{\cal P}\hat{\cal I})^2 = -1$, hence ${\cal H}_{\text{BdG}}(\bf k)$ belongs to 
CI$+{\cal I}$ class in the AZ$+{\cal I}$ table for classifying inversion symmetric Hamiltonians with band structure nodes proposed in Ref. \cite{AZ+I}. There are two topological charges associated with the CI class, i.e., elements of the 
first and second homotopy groups $\pi_1(M_{\text{CI}})$ and $\pi_2(M_{\text{CI}})$, where $M_{\text{CI}} = \text{U}(n)/\text{O}(n)$ is the classifying topological space relevant for CI. The $\pi_2$ monopole charge is trivial in our case, meaning that the nodal ring can shrink to a point and annihilate itself by continuously tuning the model parameters. Pertaining to the present case, we only focus on the effect of a nontrivial $\pi_1$ charge when the Fermi surface topology do not change. In the presence of the chiral symmetry $\hat{\cal S}$, the flattened Hamiltonian ${\cal H}_{\text{flat}}({\bf k})$ can be deformed into an off-diagonal form
\begin{align}
    {\cal H}_{\text{flat}}({\bf k}) = \[
        \begin{array}{cc}
             & q({\bf k})\\
            q^{\dag}({\bf k}) &
        \end{array}
    \right],
\end{align}
and the $\pi_1$ charge can be captured by the phase winding number of the $q(\bf k)$ matrix along an arbitrary closed path $S^1$ which interlocks with the nodal ring
\begin{align}
    \pi_1(M_{\text{CI}}) = \frac{i}{2\pi}\oint_{S^1} d{\bf k}\cdot \text{Tr}[q^{\dag}({\bf k})\nabla_{\bf k}q({\bf k})]   \in \mathbb{Z}.
    \label{eq:CI_charge}
\end{align}

\begin{figure}
    \includegraphics[width = 3in]{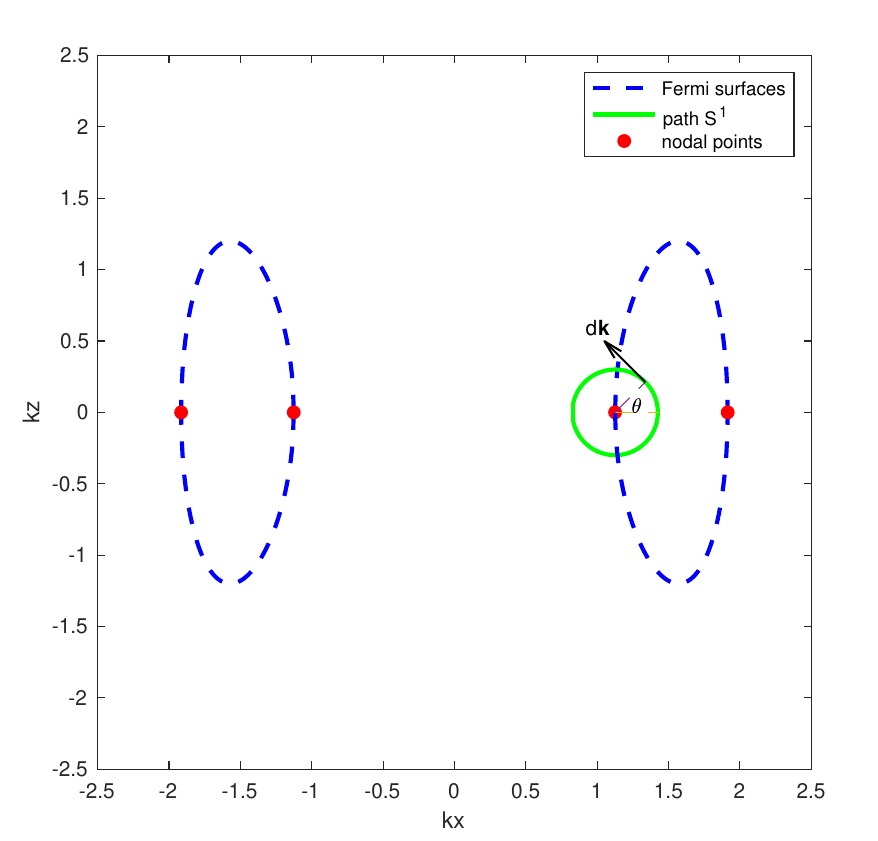}\\
    \caption{Sketch of the path $S^1$. It winds around the inner nodal ring in a counter clockwise direction.}
    \label{fig:S1}
\end{figure}

To calculate the winding number associated with each nodal ring, we follow Ref. \cite{Qi2010} to derive $q({\bf k})$. For a generic BdG Hamiltonian with chiral symmetry
\begin{align}
    {\cal H}_{\text{BdG}}({\bf k}) = \[
        \begin{array}{cc}
            H_0 ({\bf k}) & \Delta({\bf k})\\
            \Delta^{\dag}({\bf k}) & -H_0^{\text{T}} (-{\bf k})
        \end{array}
    \],
\end{align}
we can always unitarily transform it into an off-diagonal form
\begin{align}
    \tilde{{\cal H}}_{\text{BdG}} ({\bf k}) &= V{\cal H}_{\text{BdG}}({\bf k})V^{\dag} \nonumber\\
    &= \[
        \begin{array}{cc}
             & H_0 ({\bf k}) + i {\cal T}\Delta_{\bf k}^{\dag}\\
            H_0 ({\bf k}) - i {\cal T}\Delta_{\bf k}^{\dag} & 
        \end{array}
    \right],
\end{align}
where ${\cal T}$ is the unitary part of the time reversal operator and 
\begin{align}
    V = \frac{1}{\sqrt{2}}\[
        \begin{array}{cc}
            \mathbb{I} & i{\cal T}\\
            \mathbb{I} & -i{\cal T}
        \end{array}
    \].
\end{align}
Since the weak pairing $\Delta_{\bf k}$ is only turned on around the Fermi surface, the matrix elements of 
${\cal T}\Delta_{\bf k}^{\dag}$ between different bands are negligible. Therefore, we can use the Bloch states of $H_0 ({\bf k})$
to expand the off-diagonal matrix
\begin{align}
    H_0({\bf k}) + i{\cal T}\Delta^{\dag}_{\bf k} \simeq \sum_n \(\varepsilon_{n,\bf k} + i\delta_{n,\bf k}\)
    |n,{\bf k}\rangle \langle n, {\bf k}|.
\end{align}
The matrix elements $\delta_{n,\bf k}$ are 
\begin{align}
    \delta_{\pm,{\bf k}} \equiv \langle \pm,{\bf k}|{\cal T}\Delta_{\bf k}^{\dag}|\pm,{\bf k}\rangle  = \pm \frac{\Delta k_z}{k_r},
\end{align}
which are consistent with the projected gap onto the FS in Eq. \eqref{eq:projected}. Correspondingly, the off-diagonal matrix $q(\bf k)$ in the flattened Hamiltonian ${\cal H}_{\text{flat}}(\bf k)$ is given by
\begin{align}
    q({\bf k}) &= \sum_{n} e^{i\theta_{n,\bf k}} |n,{\bf k}\rangle\langle n,{\bf k}| \nonumber\\
    &=  \sum_n \frac{\varepsilon_{n,\bf k} + i\delta_{n,\bf k}}{\left|\varepsilon_{n,\bf k} + i\delta_{n,\bf k}\right|} |n,{\bf k}\rangle \langle n,{\bf k}|.
    \label{eq:qmatrix}
\end{align}
Note that $\Delta \ll \mu$, so we obtain $e^{i\theta_{-,\bf k}}\simeq -1$ and thus
\begin{align}
    q({\bf k}) \simeq e^{i\theta_{+,\bf k}}|+,{\bf k}\rangle\langle +,{\bf k}| - |-,{\bf k}\rangle\langle -,\bf k|.
\end{align}
Only the first term contains a relevant contribution to the phase winding, so we safely set $q({\bf k}) = e^{i\theta_{+,\bf k}}|+,{\bf k}\rangle\langle +,\bf k|$ and the $\pi_1$ charge is 
\begin{align}
    \pi_1(M_{\text{CI}}) = -\frac{1}{2\pi}\oint_{S^1} d{\bf k} \cdot \nabla_{\bf k}\theta_{+,\bf k} = -\frac{1}{2\pi}\theta_{+,\bf k}\Big|_{i}^{f},
    \label{eq:charge}
\end{align}
where $i$ and $f$ represents the starting and ending point of the path $S^1$. We can choose a circular path $S^1$ at $k_y = 0$, where the two nodal rings become four symmetric nodal points at $\pm k_{x1}, \pm k_{x2}$ 
with $k_{x1} \simeq \sqrt{m-\mu}$\ \ and\ \ $k_{x2} \simeq \sqrt{m+\mu}$. $S^1$ interlocks $k_{x1}$ in a counter-clockwise direction shown in Fig. \ref{fig:S1}. From Eq. \eqref{eq:qmatrix} we obtain
\begin{align}
    e^{i\theta_{+,\bf k}} = \frac{k_r - \mu + i\frac{\Delta k_z}{k_r}}{
        \[\(k_r - \mu\)^2 + \frac{\Delta^2 k_z^2}{k_r^2}\]^{1/2}}.
\end{align}
Along $S^1$, the phase $\theta_{+,\bf k}$ changes as
\begin{align}
    \theta_{+,\bf k}: \qquad \pi \to \frac{\pi}{2} \to 0 \to -\frac{\pi}{2} \to -\pi.
\end{align}
According to Eq. \eqref{eq:charge}, the topological charge of the inner superconducting nodal ring is determined as
\begin{align}
    \pi_1(M_{\text{CI}}) = -\frac{1}{2\pi} \[\theta_{+,\bf k}(f) - \theta_{+,\bf k}(i)\]  = 1.
\end{align}
Similar calculations show that the charge for the outer nodal ring is $-1$. Spin indices were suppressed throughout the calculations above, hence the winding number should be  $+2$ for the inner nodal ring and  $-2$ for the outer nodal ring after counting the spin degeneracy.  Due to the nontrivial and opposite $\pi_1$ charges carried by the pair of nodal rings, both top and bottom surface Brillouin zones  corresponding to the BdG Hamiltonian in Eq. \eqref{eq:BdG} contain flat-band Majorana zero modes enclosed by the projections of the pair of nodal rings onto the surfaces \cite{Wang_NLSC1,NRSC}.

\subsection{Ferromagnetic phase}

 \begin{table}
\begin{tabular}{c|c|c|c|c}
    \hline
    \hline
    Pairing order& Irrep & $\hat{\cal I} = \sigma_z$ & $\hat{\cal C}_{2z} = is_z$ & $\hat{\cal M}_z = i\sigma_zs_z$ \\
    \hline
    $\Delta_a\sigma_y(s_0 + s_z)$  & $B_{u}$  & $-$    & $-$ &   $+$ \\
    $\Delta_b\sigma_y(s_0 - s_z)$& $B_{u}$  &   $-$  & $-$  & $+$ \\
    $\Delta_z\sigma_ys_x$& $A_{u}$  &   $-$  & $+$  & $-$ \\
    \hline
    \hline
   \end{tabular}
\caption{List of representative pairing orders,  irreducible corepresentations, and their corresponding characters of three orbital-singlet and spin-triplet pairing channels in magnetic point group No. 8.4.27.}
\label{table:1}
\end{table}

In the presence of ferromagnetism, the symmetry of the system is lowered to the magnetic point group No. 8.4.27 \cite{BCS}. The unitary crystalline symmetries of the normal state Hamiltonian in Eq. \eqref{eq:ferro_normal} are inversion $\hat{\cal I} = \sigma_z$, two-fold rotation with respect to $z$ axis $\hat{\cal C}_{2z} = is_z$ and mirror operation with respect to $xy$ plane $\hat{\cal M}_z = i\sigma_z s_z$. As a result, the pairing orders with ${\bf d}\perp {\bf M}$ belong to the $B_u$ irrep while the order with ${\bf d}\parallel {\bf M}$ belongs to $A_u$. Their transformation properties are listed in Table. \ref{table:1}. From the energetic analysis in Sec. \ref{sec:ferro}, we have concluded that the favored pairing states are those with $\bf d$-vectors perpendicular to the magnetization axis. In this subsection, we only analyze the topological properties for ${\bf d}\perp {\bf M}$.

The $B_u$ pairing channels (${\bf d}\perp {\bf M}$) correspond to the mixed equal-spin pairing states. In order to preserve the U(1) spin rotation with respect to $z$ axis, the phases of two equal-spin pairing gaps $\Delta_a \sigma_y (s_0 + s_z)$ and $\Delta_b \sigma_y (s_0 - s_z)$ do not couple to each other from the GL free energy analysis in Sec. \ref{sec:ferro}. Therefore, we generally assume that the two pairing orders (spin are aligned as $|\uparrow\uparrow\rangle$ and $|\downarrow\downarrow\rangle$) carry arbitrary phases $\alpha$ and $\theta$. The corresponding BdG Hamiltonian is 
\begin{align}
    {\cal H}_{\text{BdG}}^{\perp}({\bf k}) = &\ (6 - t_1 - 2\cos k_x - 2\cos k_y - 2\cos k_z) \sigma_zs_0\tau_z \nonumber\\
    & + 2t_2\sin k_z \sigma_xs_0\tau_0 -\mu\sigma_0s_0\tau_z - M_z \sigma_0s_z\tau_z\nonumber\\
    & + \Delta_a \sigma_y \frac{s_0 + s_z}{2} \(\tau_x\cos\alpha + \tau_y \sin\alpha \)\nonumber\\
    & + \Delta_b \sigma_y \frac{s_0 - s_z}{2} \(\tau_x\cos\theta  + \tau_y \sin\theta \),
    \label{eq:BdG_ferro2}
\end{align}
where the inversion ($\hat{\cal I} = \sigma_z\tau_z$) and particle-hole symmetries ($\hat{\cal P} = \tau_x \cal K$) are the same as the paramagnetic phase.

\begin{figure}
	\includegraphics[width=0.8\columnwidth]{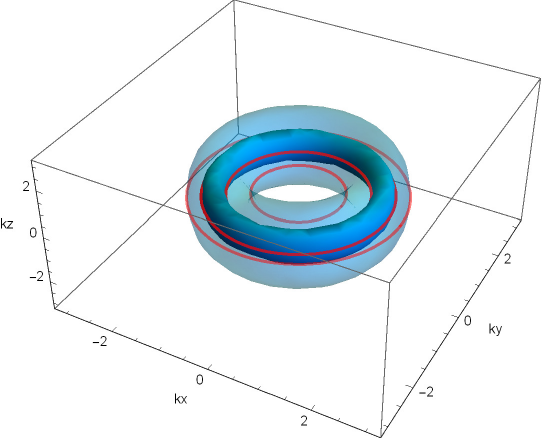}
   \caption{Superconducting nodal rings of $B_u$ pairing channels in the presence of ferromagnetism. The $\pi_1$ charges of the nodal rings are +1, +1, -1 and -1 from inside to outside.} 
\label{fig:SC_ferro}
\end{figure}

Note that while the magnetic order breaks both time-reversal symmetry  $\mathcal{T}=i\sigma_z s_y \mathcal{K}$ and spin rotation symmetry, it preserves their composite ${\cal T}' = \sigma_z\cal K$ we used to identify the CI classification. However, the pairing orders in general breaks $\mathcal{T}'$ due to the complex phases of $\Delta_{a,b}$. This can be remedied by a phase rotations in Nambu space for each of the two spins, and the modified time-reversal symmetry is
\begin{align}
{\cal T}'' = \sigma_z  \[ \frac{s_0+s_z}{2}e^{-i\alpha\tau_z} + \frac{s_0-s_z}{2}e^{-i\theta\tau_z}\]\cal K,
\end{align}
which satisfies ${\cal T}''{\cal H}_{\text{BdG}}^{\perp}({-\bf k}) {\cal T}''^{-1} = {\cal H}_{\text{BdG}}^{\perp}({\bf k})$ and $({\cal T}''{\cal I})^2 = +1$.

 We emphasize that the $\mathcal{T}''$ symmetry exists for arbitrary phases $\alpha$ and $\theta$. Consequently, the eight-band BdG Hamiltonian \eqref{eq:BdG_ferro2} belongs to class CI$+{\cal I}$ as well and it describes two decoupled superconducting orders on the spin-polarized Fermi surfaces  equivalent to the Hamiltonian in Eq. \eqref{eq:BdG}, both supporting topologically protected nodal rings on the equators shown in Fig. \ref{fig:SC_ferro}. The $\pi_1$ charges defined in Eq. \eqref{eq:CI_charge} are both +1 for two inner nodal rings and $-1$ for two outer nodal rings. Since the topological charge  is an integer quantity, nodal rings with the same sign will not pair-annihilate when the ferromagnetic order is turned off.

The nodal-loop superconductivity can be further verified by solving the BdG spectrum of Eq. \eqref{eq:BdG_ferro2} and subsequently setting $E({\bf k}) = 0$. Upon doing so, we find four nodal loops at 
\begin{align}
	k_z = 0,\quad \cos k_x + \cos k_y &= 2 - \frac{t_1}{2} \pm \frac12\sqrt{(\mu + M_z)^2 + \Delta_a^2} \nonumber\\
	\text{for} \qquad s_z = +1,\\ 
	k_z = 0,\quad \cos k_x + \cos k_y &= 2 - \frac{t_1}{2} \pm \frac12\sqrt{(\mu - M_z)^2 + \Delta_b^2} \nonumber\\
	 \text{for} \qquad s_z = -1.
\end{align}

\begin{figure*}
\centering
\includegraphics[width=1.8\columnwidth]{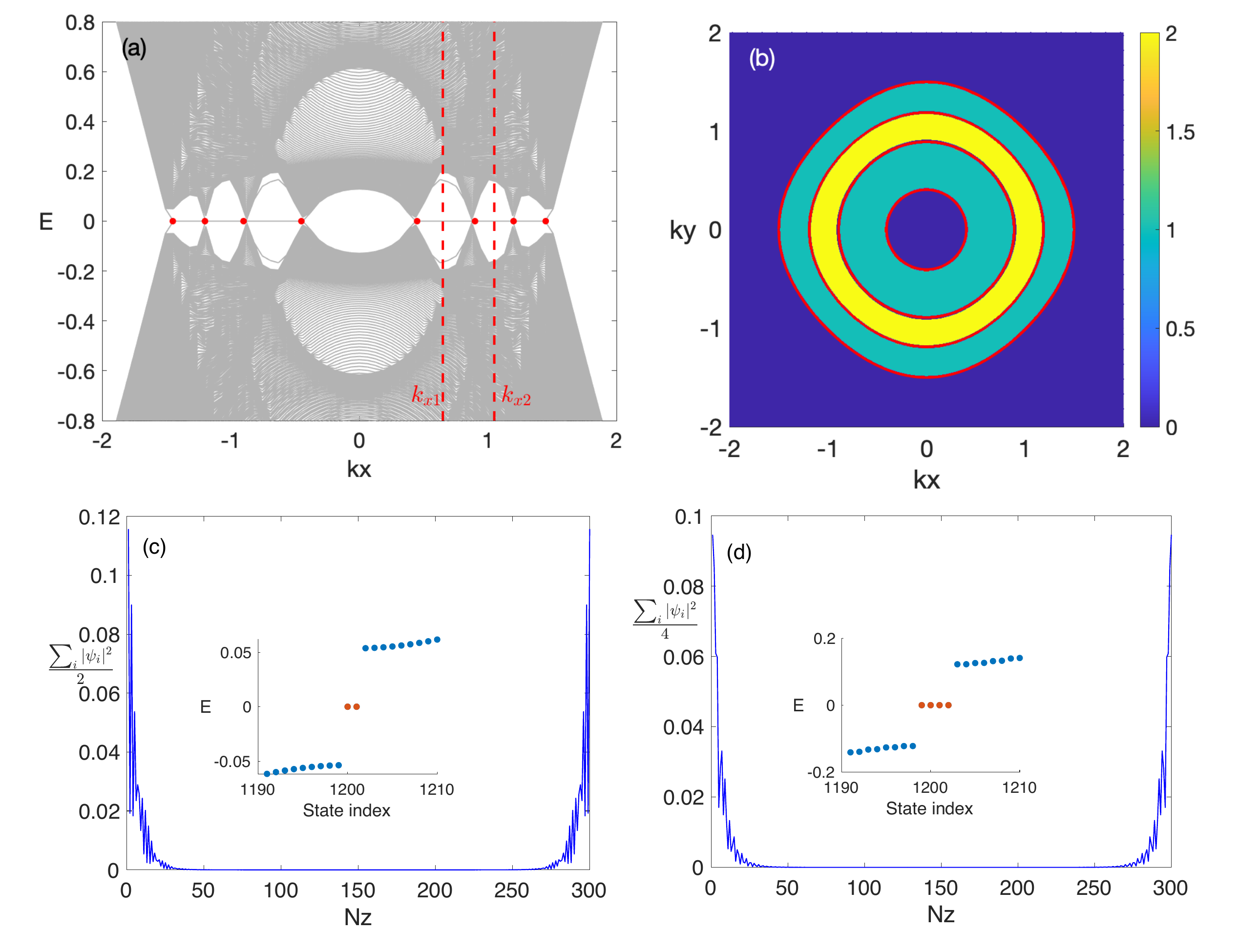}
   \caption{(a)Energy spectrum of ${\cal H}^{\perp}_\text{BdG}({\bf k})$ in Eq. \eqref{eq:BdG_ferro2} versus $k_x$ at $k_y = 0$. The open boundary condition along $z$ direction with lattice sites $N_z = 300$ is adopted in the spectrum. Red dots denote the positions of bulk superconducting nodal rings. (b)Topological regions on top (bottom) surface Brillouin Zone. The yellow region denotes $N_\text{1D} = 2$, cyan regions denote $N_\text{1D} = 1$ and blue regions denote $N_\text{1D} = 0$. Majorana zero modes and the corresponding density profiles are plotted for (c) $k_{x1} = 0.65$ and (d) $k_{x2} = 1.05$. We have set $\alpha = \theta = 0$ and $\Delta_a = \Delta_b = 0.2$. Other model parameters are set to be $\{t_1,t_2,\mu,M_z\} = \{1,0.5,0.5,0.3\}$.} 
\label{fig:surface}
\end{figure*}

For bulk-boundary correspondence, these nodal rings enclose flat-band Majorana zero modes on top and bottom surfaces of the lattice Hamiltonian in Eq. \eqref{eq:BdG_ferro2} depicted in Fig. \ref{fig:surface}. Moreover, the number of surface Majorana zero modes is determined by a $\mathbb{Z}$-valued topological invariant carried by the effective 1D Hamiltonian ${\cal H}_{\text{BdG}}^{1D}(k_z)$ by fixing $k_x$ and $k_y$ in Eq. \eqref{eq:BdG_ferro2}. For notational simplicity, we set $\alpha = \theta = 0$ and $\Delta_a = \Delta_b$ (Strictly speaking, one should set $\Delta_a>\Delta_b$ since $\Delta_a$ is the leading instability. Nevertheless, we claim that this choice only modifies the positions of the pair of nodal rings on the ``smaller FS" in Fig. \ref{fig:SC_ferro}, which does not influence the topology discussed below), yielding
\begin{align}
	{\cal H}_{\text{BdG}}^{1D}(k_z) = &\ (m - 2\cos k_z)\sigma_z s_0 \tau_z + 2t_2\sin k_z \sigma_x s_0 \tau_0 \nonumber\\
	 & - \mu \sigma_0 s_0 \tau_z - M_z \sigma_0 s_z \tau_z + \Delta_a \sigma_y s_0 \tau_x,
\end{align}
where $m \equiv 6 - t_1 - 2\cos k_x - 2\cos k_y$. The $\pi_1$ charges (winding numbers)  identified  for the 3D Hamiltonian $ {\cal H}_{\text{BdG}}^{\perp}({\bf k})$ within class CI+$\cal I$ can be interpreted as the ``difference" of the topological invariants of two fully gapped 1D subsystem ${\cal H}_{\text{BdG}}^{1D}(k_z)$ separated by the superconducting nodal rings, namely
\begin{align}
	\pi_1 (M_{\text{CI}}) = N_\text{1D}^> - N_\text{1D}^<,
\end{align}
where $\lessgtr$ denotes the region inside(outside) the corresponding nodal ring and vice versa. From deforming the loop $S^1$ (along which the $\pi_1$ charge is defined) into two straight lines that cross the 1D Brillouin Zone along $k_z$, we can determine distinct $\mathbb{Z}$-valued 1D winding numbers $N_\text{1D}$ for ${\cal H}_{\text{BdG}}^{1D}(k_z)$ (details in Appendix \ref{appen:2}). The topological invariants are found to be $N_\text{1D} = 2$ for the annulus region enclosed by two superconducting nodal rings located on the ``small" FS while $N_\text{1D} = 1$ for two other annulus regions between the ``small" and ``large" FS.

We numerically solve the energy spectrum in Fig. \hyperref[fig:surface]{\ref{fig:surface}(a)} by introducing open boundary condition along $z$, where we found flat-band Majorana zero modes enclosed by the bulk superconducting nodal-rings projected on the surface. The 1D topological invariants found above are also numerically verified in Fig. \hyperref[fig:surface]{\ref{fig:surface}(c)} and \hyperref[fig:surface]{\ref{fig:surface}(d)}. A schematic picture of the topological regions on the top and bottom surface Brillouin Zones is illustrated in Fig. \hyperref[fig:surface]{\ref{fig:surface}(b)}. We note that the surface flat-band Majorana zero modes discussed in our work are different from the Weyl-loop superconducting phases in Refs. \cite{Wang_NLSC1,NRSC}. In our Dirac-loop case, there is an extra $N_\text{1D} = 2$ region with two surface Majorana zero modes coming from two spin-polarized sectors. While the $N_\text{1D} = 2$ region also appears in a Weyl-loop superconducting model in Ref. \cite{NRSC}, the origin there is different from ours since the spin degrees of freedom are not considered in Ref. \cite{NRSC}.

We briefly comment on the case for $A_u$ pairing symmetry (${\bf d}\parallel {\bf M}$) in Appendix \ref{appen:3}, where we find two toriodal Bogoliubov Fermi surfaces that are topologically unstable. This pairing order explicitly breaks time-reversal symmetry and the system belongs to C+$\cal{I}$ class in Ref. \cite{AZ+I}. This class lacks a nontrivial $\pi_0$ topological charge, which is necessary to stabilize a nodal surface in 3D BdG spectrum. Due to the larger gapless regions in the BdG spectrum, the $A_u$ channel is suppressed, consistent with the energetic analysis  in Sec. \ref{sec:ferro}.

\section{Summary}
\label{sec:conclusion}
In this work, we analyzed the energetic and topological properties of nodal superconductivity induced by ferromagnetic spin fluctuations or Hund's coupling in Dirac-loop-type nodal-line semimetals. The favored Cooper pairing channels are found to be momentum-independent, orbital-singlet and spin-triplet, which belongs to the $B_{1u}$ representation of the point group $D_{2h}$. In the weak-pairing regime, we calculated the critical temperatures in the paramagnetic phase. From the pseudo-spin textures on the torus Fermi surface, three spin-triplet pairing channels all exhibit a pair of nodal rings, which are topologically protected by a $\mathbb{Z}$-valued charge $\nu = \pm 2$ within class CI+$\cal I$ from the AZ+$\cal I$ table \cite{AZ+I}.

In the presence of a ferromagnetic order, the symmetry of the system is lowered to the magnetic point group No. 8.4.27. We analyze Ginzburg-Landau free energy of the system which captures the interplay between spin-triplet superconductivity and ferromagnetism. The leading pairing state is found to carry a relative phase $\pi/2$ between $d_x$ and $d_y$ components, i.e., ${\bf d} \propto (1,i,0)$. Upon further lowering the temperature, a sub-leading channel with $|\downarrow\downarrow\rangle$ spin is  favored. These two pairing orders correspond to the pairing of the two split FS's with opposite spins. We show that the BdG Hamiltonian belongs to class CI+$\cal I$ since it still preserves a ``modified" time-reversal symmetry which squares to +1. Therefore, the four-fold degenerate superconducting nodal rings found from the paramagnetic phase are split into two pairs, and the robustness of nodal rings can be characterized by an integer-valued topological invariant $\nu = \pm 1$. Furthermore, the $\pi_1$ charges (winding numbers)  identified within class CI+$\cal I$ can be interpreted as the ``difference" of the topological invariants of two fully gapped 1D subsystem separated by the superconducting nodal rings. This is demonstrated in the energy spectrum with open boundary condition along $z$ direction, where we find that the bulk superconducting nodal rings enclose flat-band Majorana zero modes with $N_\text{1D} = 2$ and $N_\text{1D} = 1$ on the top and bottom surface Brillouin Zones.
For ${\bf d}\parallel \bf M$, the BdG quasiparticle spectrum hosts nodal surfaces in class C+$\cal I$ which are not topologically protected. This pairing channel is energetically disfavored because of the large gapless surfaces in the BdG spectrum.

The nodal-ring superconductivity discussed in our theory can be applied to either paramagnetic nodal-line systems (mediated by Hund's coupling) or ferromagnetic nodal-line materials (mediated by ferromagnetic spin fluctuations).  For the latter case,  superconductivity may appear close to onset of ferromagnetism, although experimentally the bulk superconductivity is yet to be discovered. The superconducting phase should naturally host nodal rings inherited from the normal state.  While our analysis is based on a simplified Hamiltonian, we expect the conclusions to hold for realistic materials so long as the corresponding symmetries are the same. Candidate Dirac loop materials such as Ca$_3$P$_2$ \cite{Chan2016}, Cu$_3$N \cite{Dirac_loop}, CaAgP and CaAgAs \cite{CaAgP} might be promising platforms, while future studies are necessary to investigate  potential superconducting phases in these and related materials.

\acknowledgements
We acknowledge the support from startup funds at University of Florida and National Science Foundation (NSF) under Award number DMR-2045781.

\appendix

\section{Fierz identity}
\label{appen:1}
\subsection{Proof}
\label{appen:1.1}
Fierz identities are reordering relations for four-fermion interactions: for two $n\times n$ matrices $M, N$ and $\psi_i$ as $n$-component fermionic 
annihilation operators, there exist matrices $M^{\prime}, N^{\prime}$ such that 
\begin{align}
    \psi_{1}^{\dag} M \psi_2 \psi_3^{\dag} N \psi_4 = \psi_1^{\dag} M^{\prime} \(\psi_3^{\dag}\)^{\text{T}} \psi_4^{\text{T}} N^{\prime} \psi_2,
    \label{eq:A1}
\end{align}
to prove the relation in Eq. \eqref{eq:A1}, we first choose a set of orthogonal basis of $n\times n$ Hilbert space $\{Q_a\}, a = 1, 2, 3, \dots, n^2$, which satisfies
\begin{align}
    \text{Tr}\(Q_a Q_b^{\dag}\) = n\delta_{ab},
\end{align}
an arbitrary $n\times n$ matrix $M$ can be expanded as
\begin{align}
    M = \sum_a M_a Q_a, \quad M_a = \frac{1}{n} \text{Tr}\(MQ_a^{\dag}\).
\end{align}

Note that 
\begin{align}
    \psi_1^{\dag} M \psi_2 \psi_3^{\dag} N \psi_4 & =  \psi_{1i}^{\dag} M_{ij} \psi_{2j} \psi_{3k}^{\dag} N_{kl} \psi_{4l} \nonumber\\
    & =  M_{ij} N_{kl} \psi_{1i}^{\dag} \psi_{3k}^{\dag} \psi_{4l} \psi_{2j},    
    \label{eq:A4}
\end{align}
where identical indices are summed over. We can further expand 
\begin{align}
    M_{ij} N_{kl} = \sum_{ab}C_{ab} (Q_a)_{ik}(Q_b^{\dag})_{lj},
    \label{eq:A5}
\end{align}
the coefficient $C_{ab}$ can be determined by multiplying both sides of Eq.\eqref{eq:A5} with $(Q_c^{\dag})_{\lambda i}$ and $(Q_d)_{\rho l}$ and sum over $i, l$
\begin{align}
    (Q_c^{\dag}M)_{\lambda j} (Q_d N^{\text{T}})_{\rho k} = \sum_{ab}C_{ab} (Q_{c}^{\dag}Q_a)_{\lambda k} (Q_d Q_b^{\dag})_{\rho j},
\end{align}
set $\lambda = k$ and $\rho = j$ and sum over $\lambda, k$, we obtain
\begin{align}
    \text{Tr} \(Q_c^{\dag} M Q_{d} N^{\text{T}}\) = n^2 \sum_{ab} C_{ab} \delta_{ac}\delta_{bd} = n^2 C_{cd},
\end{align}
therefore, the coefficient $C_{ab}$ is given by
\begin{align}
    C_{ab} = \frac{1}{n^2} \text{Tr} \(Q_a^{\dag} M Q_b N^{\text{T}}\),
\end{align}
yielding
\begin{align}
    M_{ij}N_{kl} = \frac{1}{n^2}\sum_{ab} \text{Tr} \(Q_a^{\dag} M Q_b N^{\text{T}}\)(Q_a)_{ik}(Q_b^{\dag})_{lj},
    \label{eq:A9}
\end{align}
combine Eq. \eqref{eq:A4} and Eq. \eqref{eq:A9}, we obtain
\begin{align}
    &\psi_{1}^{\dag} M \psi_2 \psi_3^{\dag} N \psi_4 \nonumber\\
    &= \frac{1}{n^2} \sum_{ab} \text{Tr} \(Q_a^{\dag} M Q_b N^{\text{T}}\) \psi_1^{\dag} Q_a \left(\psi_3^{\dag}\right)^{\text{T}} \psi_4^{\text{T}} Q_b^{\dag} \psi_2,
    \label{eq:fierz}
\end{align}
hence we have proved the Fierz identity.

\begin{widetext}
\subsection{Decompositions from ferromagnetic fluctuations}
\label{appen:1.2}
The four-fermion interaction in Eq. \eqref{eq:pairing} in the main text can be decomposed into orbital and spin channels independently. Set  $n = 2$ and choose 
\begin{align}
&\sigma_0i\sigma_y,\quad \sigma_xi\sigma_y,\quad \sigma_yi\sigma_y,\quad \sigma_zi\sigma_y\qquad \text{for orbital subspace}, \nonumber\\
&s_0is_y,\quad s_xis_y,\quad s_yis_y,\quad s_zis_y\qquad  \text{for spin subspace}. \nonumber
\end{align}
For orbital decompositions, we treat $c^{\dag}({\bf p})$ as a two component vector in orbital space and set $M = N = \sigma_0$ in Eq. \eqref{eq:fierz} to obtain
\begin{align}
    c^{\dag}({\bf p})\sigma_0\vec{s}\ c({\bf k})\cdot  c^{\dag}(-{\bf p})\sigma_0\vec{s}\ c(-{\bf k})
    = \frac{1}{2}\sum_{a = 0,x, y, z} c^{\dag}({\bf p})(\sigma_a i\sigma_y)\otimes\vec{s} \[c^{\dag}(-{\bf p})\]^{\text{T}}\cdot \[c(-{\bf k})\]^{\text{T}}\(\sigma_a i\sigma_y\)^{\dag} \otimes \vec{s}\ c({\bf k}).
    \label{eq:orbit}
\end{align}
For spin decompositions, we treat $c^{\dag}({\bf p})$ as a two component vector in spin space and set $M = N = s_x, s_y, s_z$ in Eq. \eqref{eq:fierz} separately, yielding
\begin{align}
    c^{\dag}({\bf p})\sigma_0s_x c({\bf k}) c^{\dag}(-{\bf p})\sigma_0s_x c(-{\bf k}) 
    &= \frac{1}{2}\(-b^{\dag}_{{\bf p},0} b_{{\bf k},0} - b^{\dag}_{{\bf p},x} b_{{\bf k},x} + b^{\dag}_{{\bf p},y} b_{{\bf k},y} + b^{\dag}_{{\bf p},z} b_{{\bf k},z}\),  \\
    c^{\dag}({\bf p})\sigma_0s_y c({\bf k}) c^{\dag}(-{\bf p})\sigma_0s_y c(-{\bf k}) 
    &= \frac{1}{2}\(-b^{\dag}_{{\bf p},0} b_{{\bf k},0} + b^{\dag}_{{\bf p},x} b_{{\bf k},x} - b^{\dag}_{{\bf p},y} b_{{\bf k},y} + b^{\dag}_{{\bf p},z} b_{{\bf k},z}\),  \\
    c^{\dag}({\bf p})\sigma_0s_z c({\bf k}) c^{\dag}(-{\bf p})\sigma_0s_z c(-{\bf k}) 
    &= \frac{1}{2}\(-b^{\dag}_{{\bf p},0} b_{{\bf k},0} + b^{\dag}_{{\bf p},x} b_{{\bf k},x} + b^{\dag}_{{\bf p},y} b_{{\bf k},y} - b^{\dag}_{{\bf p},z} b_{{\bf k},z}\),  
\end{align}
where we have defined $b^{\dag}_{{\bf p},a} = c^{\dag}({\bf p})\sigma_0\otimes (s_ais_y) \[c^{\dag}(-{\bf p})\]^{\text{T}}$. After summing over three equations above, we find
\begin{align}
    c^{\dag}({\bf p})\sigma_0\vec{s}\ c({\bf k})\cdot c^{\dag}(-{\bf p})\sigma_0\vec{s}\ c(-{\bf k})
    = \frac{1}{2}\(-3b^{\dag}_{{\bf p},0} b_{{\bf k},0} + b^{\dag}_{{\bf p},x} b_{{\bf k},x} + b^{\dag}_{{\bf p},y} b_{{\bf k},y} + b^{\dag}_{{\bf p},z} b_{{\bf k},z}\).
    \label{eq:spin}
\end{align}
Combine the results of Eq. \eqref{eq:orbit} and Eq. \eqref{eq:spin}, we obtain
\begin{align}
    c^{\dag}({\bf p})\sigma_0\vec{s}\ c({\bf k}) \cdot c^{\dag}(-{\bf p})\sigma_0\vec{s}\ c(-{\bf k})
    = \frac{1}{4}\sum_{\substack{a = 0,x,y,z\\ b = x,y,z}}c^{\dag}({\bf p})\sigma_ai\sigma_y\otimes s_bis_y\[c^{\dag}(-{\bf p})\]^{\text{T}}\times \[c(-{\bf k})\]^{\text{T}}(\sigma_ai\sigma_y)^{\dag}\otimes (s_bis_y)^{\dag} c({\bf k}),
\end{align}
where the first term in Eq. \eqref{eq:spin} is neglected since only attractive channels favor superconductivity.

\subsection{Decompositions from Hund's coupling}
\label{appen:1.3}
The Hund's coupling interaction is given in Eq. \eqref{eq:hunds} and we find that only orbital space decomposition is required. Accordingly, we set $M = (\sigma_0 + \sigma_z)/2$ and $N = (\sigma_0 - \sigma_z)/2$ in Eq. \eqref{eq:fierz}. The $2\times 2$ orbital space is spanned $\{Q_a\} = \{\sigma_ai\sigma_y\}$ as well. After a complete analysis, we obtain 
\begin{align}
	 \text{Tr} \(Q_a^{\dag} M Q_b N^{\text{T}}\) = \left\{\begin{array}{l}
	 1,\qquad  \text{if\quad} \{a,b\} = \{0,0\}, \{0,z\}, \{z,0\}\ \text{or}\ \{z,z\}, \\
	 0,\qquad \text{otherwise}.
	 \end{array}\right.
\end{align}
Therefore, we obtain 
\begin{align}
    c^{\dag}({\bf p})\frac{(\sigma_0 + \sigma_z)}{2}\otimes\vec{s}\ c({\bf k}) \cdot c^{\dag}(-{\bf p})\frac{(\sigma_0 - \sigma_z)}{2}\otimes\vec{s}\ c(-{\bf k}) = \frac{1}{4}\sum_{\{i,j\}} c^{\dag}({\bf p})\sigma_i i\sigma_y \otimes \vec{s} \[c^{\dag}(-{\bf p})\]^{\text{T}} \cdot  \[c(-{\bf k})\]^{\text{T}}\(\sigma_j i\sigma_y\)^{\dag}\otimes \vec{s}\ c({\bf k}),
\end{align}
where $\{i,j\}$ is chosen from $\{0,0\},\{0,z\},\{z,0\},\{z,z\}$. The spin decomposition is the same as Eq. \eqref{eq:spin}, hence we obtain the Fierz identity relation for Hund's coupling as follows, 
\begin{align}
    c^{\dag}({\bf p})\frac{(\sigma_0 + \sigma_z)}{2}\otimes\vec{s}\ c({\bf k}) \cdot c^{\dag}(-{\bf p})\frac{(\sigma_0 - \sigma_z)}{2}\otimes\vec{s}\ c(-{\bf k}) 
        = &\ \frac{1}{8}\sum_{\substack{\{i,j\}\\ b = x,y,z}}c^{\dag}({\bf p})\sigma_i i\sigma_y\otimes s_bis_y\[c^{\dag}(-{\bf p})\]^{\text{T}}\nonumber\\
         &\hspace{1.0in} \times \[c(-{\bf k})\]^{\text{T}}(\sigma_j i\sigma_y)^{\dag}\otimes (s_bis_y)^{\dag} c({\bf k}).
\end{align}

\begin{figure}
	\subfloat[$i\gamma$]{\includegraphics[width=0.33\columnwidth]{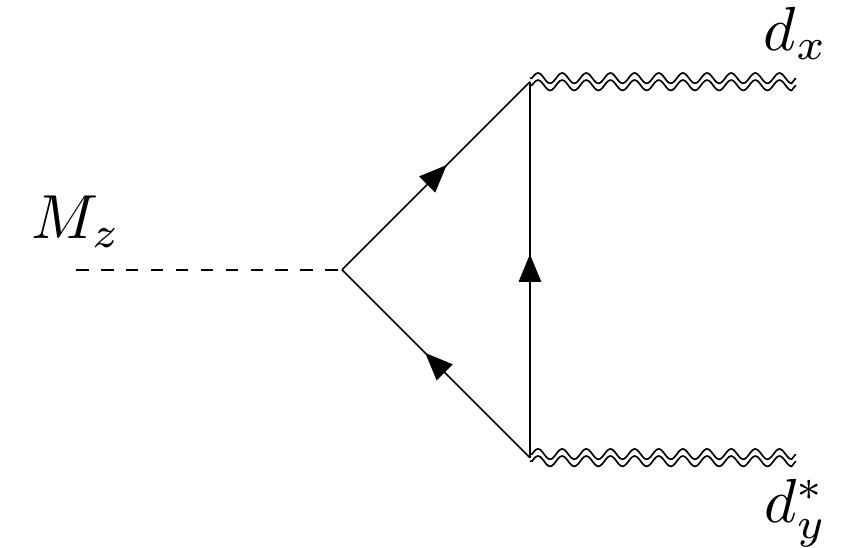}}
 \subfloat[$\beta_a$]{\includegraphics[width=0.33\columnwidth]{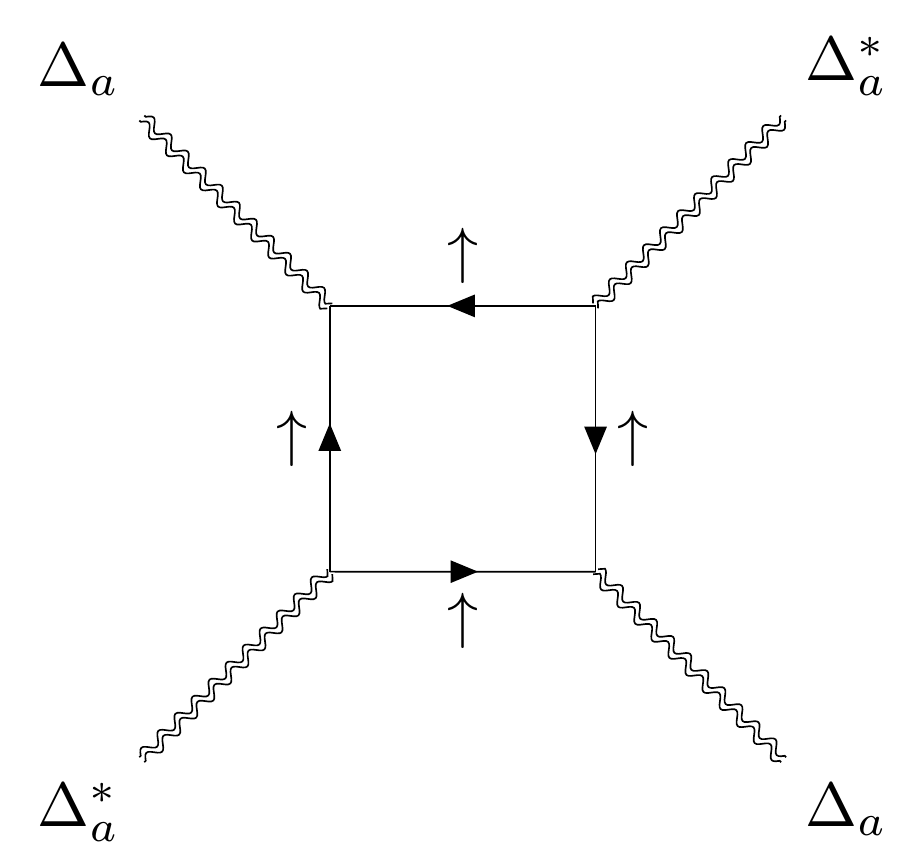}}
 \subfloat[$4\tilde{\beta}$]{\includegraphics[width=0.33\columnwidth]{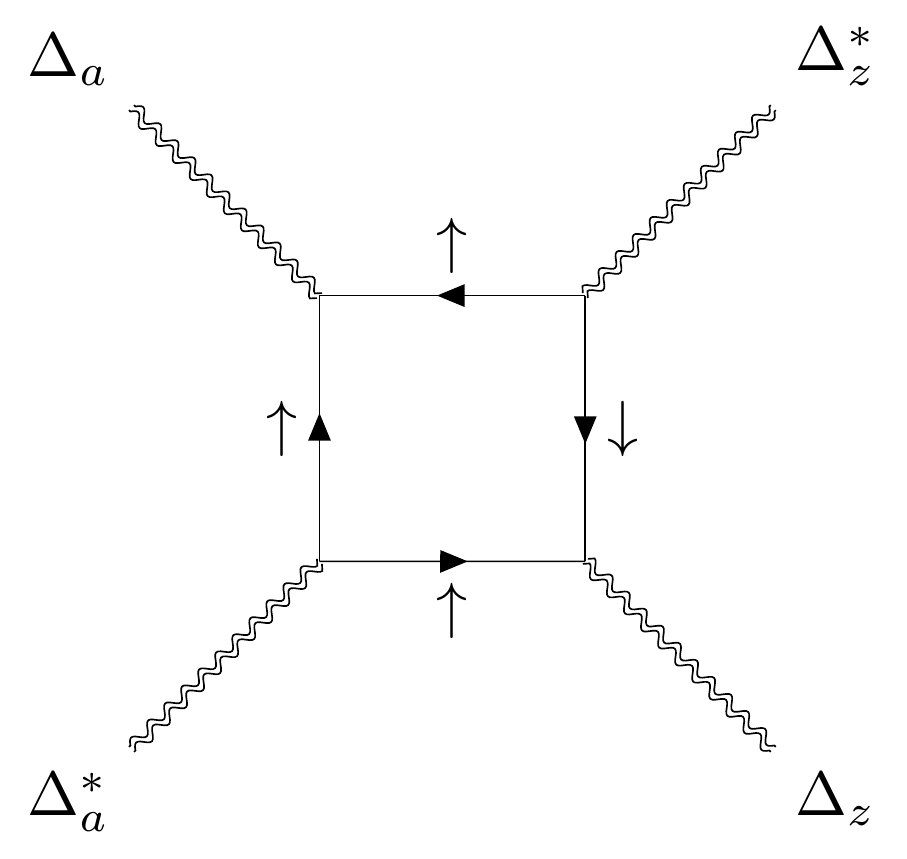}}
   \caption{Feynman diagrams relevant for the coefficients $\gamma,\beta_a,\tilde{\beta}$ in the GL free energy. Expressions of vertices are not shown in the figure. The arrows in (b) and (c) denote the fermion spin sectors.}
   \label{fig:GL}
\end{figure}

\section{Evaluations of $\gamma,\beta_a,\tilde{\beta}$ in the Ginzburg-Landau free energy}
\label{appen:diagram}

From Fig. \hyperref[fig:GL]{\ref{fig:GL}(a)}, we obtain 
\begin{align}
i\gamma &= -T\sum_k \text{Tr}\[s_z G_0(k) (-i\sigma_ys_z) G_0^{\text{T}}(-k)(-\sigma_y)^{\dag}G_0(k)\] \nonumber\\
&  = -4iT \sum_{{\bf k},n}  \frac{\[(i\omega_n+\mu)^2 + k_r^2\](-i\omega_n + \mu) }{\[(i\omega_n+\mu)^2 -k_r^2\]^2\[(-i\omega_n+\mu)^2 - k_r^2\]} \nonumber\\
&= -4i \sum_{\bf k}\int_{C} \frac{dz}{2\pi i} f(z)\frac{\[(z+\mu)^2 + k_r^2\](-z+\mu)}{\[(z+\mu)^2 - k_r^2\]^2\[(-z+\mu)^2 - k_r^2\]}  \nonumber\\
&  \simeq 4iN(0) \int d\epsilon\frac{f'(\epsilon)}{8\mu}\nonumber\\
 &= -\frac{iN(0)}{\mu}.
\label{eq:gamma}
\end{align}

The coefficients $\beta_a$ and $\tilde{\beta}$ can be distinguished by the spin vertices and symmetry factors in the Feynman diagrams shown in Fig. \hyperref[fig:GL]{\ref{fig:GL}(b)} and \hyperref[fig:GL]{\ref{fig:GL}(c)}, which are
\begin{align}
	\beta_a = &\ \frac{\beta}{4} \text{Tr}\[(-s_+)(-s_+)^{\dag}(-s_+)(-s_+)^{\dag}\] = \frac{\beta}{4}, \\
	4\tilde{\beta} = &\ \beta\text{Tr}\[(-s_+)(s_x)^{\dag}(s_x)(-s_+)^{\dag}\] = \beta,
\end{align}
where $s_+ = (s_0 + s_z)/2$ and 
\begingroup
\allowdisplaybreaks
\begin{align}
	\beta & = T\sum_k \text{Tr}\big[(i\sigma_y)G_0^\text{T}(-k)(i\sigma_y)^{\dag}G_0(k)(i\sigma_y)G_0^\text{T}(-k)\nonumber\\
	&\hspace{2.3in} \times(i\sigma_y)^{\dag}G_0(k)\big] \nonumber\\
&  = 2T\sum_{{\bf k},n}  \frac{(\omega_n^2+\mu^2-k_p^2+k_z^2)^2 - 4\omega_n^2 k_p^2 +4k_z^2(\mu^2-k_p^2)}{\[(i\omega_n+\mu)^2 -k_r^2\]^2\[(-i\omega_n+\mu)^2 - k_r^2\]^2} \nonumber\\
&  \simeq 2N(0)T\sum_n \int d\epsilon  \frac{\omega_n^4 + 4\mu^4}{(\omega_n^2 + 4\mu^2)^2(\omega_n^2 + \epsilon^2)^2} \nonumber\\
&  \simeq 4\pi N(0)T \int_{\pi T}^{\infty}   \frac{d\omega}{2\pi T} \frac{\omega^4 + 4\mu^4}{\omega^3(\omega^2 + 4\mu^2)^2} \nonumber\\
& \simeq \frac{N(0)}{4\pi^2 T^2},
\end{align}
\endgroup
where the condition $\pi T \ll \mu$ is utilized throughout the calculation.

\end{widetext}

\section{Topological invariant of ${\cal H}_{\text{BdG}}^{1D}(k_z)$}
\label{appen:2}
The topological invariant of ${\cal H}_{\text{BdG}}^{1D}(k_z)$ can be evaluated from the method discussed in Ref. \cite{BDI,Qin2022,Lapp2023}, 

\begin{align}
	N_\text{1D} = -\frac{i}{\pi}\int_{k_z = 0}^{k_z = \pi} \frac{dz(k_z)}{z(k_z)}, 
	\label{eq:BDI}
\end{align}
where $z(k_z) \equiv e^{i\theta(k_z)} = \text{det}Q(k_z)/|\text{det}Q(k_z)|$ and $Q(k_z)$ is the off-diagonal block matrix of $\tilde{{\cal H}}_{\text{BdG}}^{1D}(k_z)$, which is the original 1D BdG Hamiltonian under the rotation in Nambu space ($U = e^{-i\frac{\pi}{4}\tau_y}$)
\begin{align}
	\tilde{{\cal H}}_{\text{BdG}}^{1D}(k_z)  = U{\cal H}_{\text{BdG}}^{1D}(k_z) U^{\dag} = \[\begin{array}{cc}
     & Q(k_z)\\
    Q^{\dag}(k_z) & \end{array} \].
\end{align}
Via direct calculations, the expression of $Q(k_z)$ is 
\begin{align}
	Q(k_z) = &\ (m - 2\cos k_z)\eta_z + (2t_2\sin k_z + i\Delta_a)\eta_x \nonumber\\
	& - (\mu + M_z s_z)\eta_0,
\end{align}
where $\eta_i$ denotes the Pauli matrices in the $2\times 2$ Hilbert space spanned by $\langle\sigma_z\tau_z = +1|\sigma_i\tau_j|\sigma_z\tau_z = -1\rangle$. Since $Q(k_z)$ is diagonal in the spin space, the topological invariant in Eq. \eqref{eq:BDI} can be written as $N_\text{1D} = N_\text{1D}^+ + N_\text{1D}^-$, where $\pm$ denotes the eigenvalues of $s_z$. We further note that 
\begin{align}
	\text{det}Q_{\pm}(k_z) = &\ \Big[(\mu\pm M_z)^2 + \Delta_a^2 - (m-2\cos k_z)^2 \nonumber\\
	          &   - 4t_2^2\sin^2 k_z - 4i\Delta_a t_2\sin k_z\Big].
\end{align}
From the definition in Eq. \eqref{eq:BDI}, we obtain 
\begin{align}
	N_\text{1D}^+ = \Bigg\{\begin{array}{c}
	1 \quad \text{if}\quad \text{det}Q_+(0)\text{det}Q_+(\pi) < 0 \\
	0 \quad \text{if}\quad \text{det}Q_+(0)\text{det}Q_+(\pi) > 0
	\end{array}.
\end{align}
A similar conclusion holds for $N_\text{1D}^-$. For the nodal-line system considered throughout this work, the Hamiltonian at $k_z = \pi$ always yields $\text{det}Q_{\pm}(\pi) < 0$. After summing over the spin indices, $N_\text{1D}$ can be generally determined as 
\begin{align}
N_\text{1D} = \left\{\begin{array}{cl}
	2 \quad & \text{if}\quad (m-2)^2 < (\mu - M_z)^2 + \Delta_a^2, \\
	1 \quad &\text{if}\quad (\mu - M_z)^2 + \Delta_a^2 < (m-2)^2 \\
	             &\text{or}\quad  (m-2)^2 < (\mu + M_z)^2 + \Delta_a^2,  \\
	0 \quad &\text{otherwise}.
	\end{array}\right. 
\end{align}
Since $m \equiv 6 - t_1 - 2\cos k_x -2\cos k_y$ by definition, the first inequality above represents the annulus region enclosed by the two nodal rings on ``small" FS and the second inequality denotes two other annulus regions between the ``small" and ``large" FS.

\section{$A_u$ pairing channel}
\label{appen:3}

\begin{figure}
	\includegraphics[width=0.8\columnwidth]{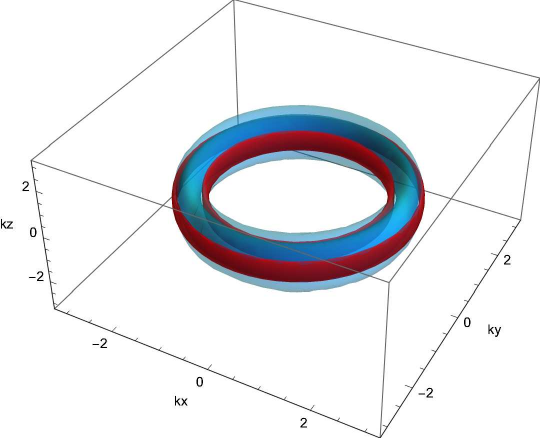}
   \caption{A pair of toroidal nodal surfaces appear on the equators in the $A_u$ pairing channel, which can be gapped out by symmetry preserved perturbations.} 
\label{fig:BGFS}
\end{figure}

When $\bf d$-vector is parallel to the magnetization axis, the pairing function belongs to $A_u$ representation in the magnetic point group listed in Table. \ref{table:1}. In this case, the system explicitly breaks time-reversal symmetry and the BdG Hamiltonian is
\begin{align}
    {\cal H}_{\text{BdG}}^{\parallel}({\bf k}) = &\ (6 - t_1 - 2\cos k_x - 2\cos k_y - 2\cos k_z) \sigma_zs_0\tau_z \nonumber\\
    & + 2t_2\sin k_z \sigma_xs_0\tau_0 -\mu\sigma_0s_0\tau_z \nonumber\\
    & - M_z \sigma_0s_z\tau_z + \Delta_z\sigma_ys_x\tau_y,
     \label{eq:BdG_ferro3}
\end{align}
which only preserves inversion ($\hat{\cal I} = \sigma_z\tau_z$) and particle-hole ($\hat{\cal P} = \tau_x\cal K$) symmetries. ${\cal H}_{\text{BdG}}^{\parallel}({\bf k})$ belongs to class C+$\cal I$ because $(\hat{\cal P}\hat{\cal I})^2 = -1$ \cite{AZ+I}. The quasiparticle spectrum $E(\bf k)$ contains nodal surfaces that satisfy 
\begin{align}
&\[f^2({\bf k}) + 4t_2^2\sin^2 k_z - \mu^2 - \Delta_z^2 -M_z^2\]^2 \nonumber\\
 &\qquad \qquad \quad + 16t_2^2\Delta_z^2\sin^2 k_z = 4\(\mu^2 + \Delta_z^2\)M_z^2,
\label{eq:BGFS}
\end{align}
where $f({\bf k}) \equiv 6 - t_1 - 2\cos k_x-2\cos k_y-2\cos k_z$. Eq. \eqref{eq:BGFS} describes a pair of toroidal Bogoliubov Fermi surfaces plotted in Fig.  \ref{fig:BGFS}. Different from Refs. \cite{BGFS1,BGFS2}, the nodal surfaces in Eq. \eqref{eq:BGFS} are not topologically protected because of the absence of Pfaffian-like topological charges in class C+$\cal I$ \cite{AZ+I}. As a simple proof, there are two $p$-wave pairing terms $\sin k_x \sigma_z\tau_y$ and $\sin k_x\sigma_z s_z\tau_x$ which preserve both $\hat{\cal P}$ and $\hat{\cal I}$ but gap out the nodal surfaces from Eq. \eqref{eq:BdG_ferro3}.

\bibliographystyle{apsrev4-1}

\bibliography{ref}

\end{document}